\documentclass[fleqn,usenatbib]{mnras}

\usepackage{newtxtext,newtxmath}

\usepackage[T1]{fontenc}

\DeclareRobustCommand{\VAN}[3]{#2}
\let\VANthebibliography\thebibliography
\def\thebibliography{\DeclareRobustCommand{\VAN}[3]{##3}\VANthebibliography}

\usepackage{graphicx}	
\usepackage{amsmath}	
\usepackage{pdflscape}

\newcommand{\kms}{\,km\,s$^{-1}$\,}

\newcommand{\feh}{[\rm Fe/H]}

\title[IFU spectroscopy of dwarf galaxies ]{The Delegate Survey: KOALA/AAOmega-IFU spectroscopy of three nearby dwarf galaxies}

\author[Kyungdo Ko]{
Kyungdo Ko,$^{1}$\thanks{E-mail: kyungdo.ko@anu.edu.au}
Helmut Jerjen,$^{1}$
Sarah M. Sweet$^{2}$
Ángel R. López-Sánchez$^{3}$
\newauthor
Simon Deeley$^{2}$
Caroline Foster$^{4}$ 
Holger Baumgardt$^{2}$
Pratyush Kumar Das$^2$
Adebusola Alabi$^{5}$
\\
$^{1}$ Research School of Astronomy and Astrophysics, The Australian National University, Canberra, ACT 2611, Australia\\
$^{2}$ School of Mathematics and Physics, University of Queensland, St Lucia,
Brisbane, QLD 4072\\
$^{3}$ Department of Physics and Astronomy, Macquarie University, Sydney, NSW 2109\\
$^{4}$ School of Physics, University of New South Wales, Sydney, NSW 2052\\
$^{5}$ University of California Observatories, 1156 High Str, Santa Cruz, CA 95064, USA
}

\date{Accepted XXX. Received YYY; in original form ZZZ}

\pubyear{\the\year{}}

\begin{document}
\label{firstpage}
\pagerange{\pageref{firstpage}--\pageref{lastpage}}
\maketitle

\begin{abstract}
Assessing the cosmological typicality of the Local Group is crucial for establishing whether its dwarf galaxy population is representative. To this end, the Delegate Survey targets dwarf galaxies in Local Group analogues for systematic comparison. As a pilot study for the Delegate Survey, we obtained integral-field spectra of three dwarf galaxies  — GAMA~79098, GAMA~569709, and HIPASS~J1159-19~S2 — in two nearby galaxy groups using the KOALA instrument on the AAT. The galaxies were mapped at a spatial sampling of $1\farcs25$\,spaxel$^{-1}$, providing two-dimensional spectral coverage of both the stellar and the ionised gas components. Using continuum and emission-line maps, we investigate the ionisation structure, star-forming properties, metallicity, star formation histories, and velocity fields of each system. 
GAMA~79098 exhibits properties closely analogous to the SMC, with ongoing star formation superimposed on an old stellar population and moderate rotational support. GAMA~569709 resembles transition-type Local Group (LG) dwarfs such as NGC~147 and NGC~185, showing predominantly an old stellar population. HIPASS~J1159-19~S2 is dominated by young stars and intense star-forming complexes, consistent with a dynamically young, gas-rich galaxy group environment. The two GAMA dwarfs follow the LG stellar mass–metallicity relation within uncertainties, while HIPASS~J1159-19~S2 appears $\approx 0.5$\,dex more metal-rich for its stellar mass. Overall, the structural, chemical, and kinematic properties of these dwarfs overlap substantially with analogous LG dwarfs.

\end{abstract}

\begin{keywords}
galaxies: groups: individual: NGC5713/19, HIPASS J1159-19 -- galaxies: dwarf -- galaxies: star formation  -- galaxies: ISM -- galaxies: kinematics and dynamics  -- galaxies: abundances 
\end{keywords}



\section{Introduction}

Most galaxies reside in groups or clusters, with galaxy groups being the most common environment in the Universe \citep{Lietzen_2012}. Groups typically host a few luminous $L_*$ galaxies accompanied by an entourage of dwarf galaxies and are embedded in dark matter haloes of masses $10^{12}$--$10^{13}\,M_\odot$. They are characterised by modest velocity dispersions of a few hundred \kms. The less severe environmental conditions and lower velocity dispersions of galaxy groups, compared with high-density clusters, promote galaxy interactions and mergers, making groups ideal laboratories for studying galaxy evolution and halo assembly histories \citep{Mamon_1999,Mihos_2003}. Low-mass dwarf galaxies are particularly susceptible to these processes because their shallow gravitational potentials make them more strongly affected by stellar feedback and environmental perturbations \citep{Wechsler_2018}. However, their intrinsically small sizes and low surface brightnesses make detailed studies beyond the nearby Universe observationally challenging. Consequently, the most detailed constraints on dwarf-galaxy stellar populations, chemical evolution, internal dynamics, and star formation histories still largely originate from the Local Group (LG) \citep{Mateo_1998,McConnachie_2012,Simon_2019,Pace_2025}.

The LG, dominated by the Milky Way (MW) and M31, hosts a rich population of dwarf galaxies spanning a wide range of luminosities, morphologies, and evolutionary states \citep{Grebel_2000,McConnachie_2012}. Wide-field imaging and spectroscopic surveys conducted over the past decade have substantially expanded the known population, particularly at the ultra-faint end, while enabling increasingly detailed studies of their stellar populations, chemical abundances, and dark matter content \citep{Simon_2019}. Compilations such as the Local Volume Database now provide an extensive and continually updated collection of the structural, kinematic, chemical, and dynamical properties of dwarf galaxies in the LG and nearby volume \citep{Pace_2025}. Their proximity has enabled detailed studies that underpin key results in dwarf galaxy science, including dynamical mass estimators and the stellar mass--metallicity relation \citep{Wolf_2010,Kirby_2013}. The LG therefore provides the best-resolved benchmark for dwarf-galaxy studies, but, as a single galaxy-group environment with a particular assembly history, it cannot automatically be assumed to be representative of galaxy groups more broadly.

Considerable progress has already been made in placing the LG satellite population in a broader extragalactic context. The Satellites Around Galactic Analogs (SAGA) and Exploration of Local VolumE Satellites (ELVES) surveys have constructed systematic satellite censuses around samples of MW-like hosts, providing constraints on satellite abundance, luminosity functions, radial distributions, and quenched fractions \citep{Mao_2024,Carlsten_2022}. Deep resolved-star searches have also begun to characterise satellite populations around hosts less massive than the MW \citep{Garling_2021}. Detailed studies of the Centaurus~A system have measured its faint satellite luminosity function and investigated the spatial flattening and apparent kinematic coherence of its satellites \citep{Crnojevic_2019,Muller_2018}. Similar work on the NGC~4490 system has compared its flattened and kinematically correlated satellite distribution with cosmological simulations \citep{Pawlowski_2024}. The MATLAS survey has additionally identified a large population of low-surface-brightness dwarf candidates across low- and moderate-density environments and enabled statistical tests of dwarf abundance outside the LG \citep{Habas_2020,Kanehisa_2024}. Together, these studies demonstrate substantial host-to-host diversity and provide increasingly powerful tests of whether individual properties of the LG satellite population are cosmologically typical.

Despite this progress, observational coverage of the different internal properties of dwarf galaxies remains uneven. Existing surveys provide strong constraints on satellite numbers, luminosity functions, projected spatial distributions, and global star formation properties, but homogeneous spatially resolved spectroscopy across dwarfs in multiple, consistently selected group environments remains comparatively limited. 

Comparisons between the LG and predictions from $\Lambda$CDM simulations have further raised questions about its typicality. Cosmological simulations of galaxy formation have highlighted several small-scale tensions, including the missing-satellite problem, the too-big-to-fail problem, shallow central density profiles in dwarfs, and the existence of coherent satellite planes \citep{Kuhlen_2008,Kimmel_2014,Nelson_2015,Bullock_2017,Pawlowski_2013}. Comparisons with multiple external systems are therefore necessary to determine whether the LG lies within the expected cosmic variance.

To complement these existing satellite surveys, we initiated the Delegate\footnote{\url{https://www.delegatesurvey.science/}} Survey to obtain resolved integral-field spectroscopy of a large number ($\sim40$) of dwarf galaxies in five carefully selected Local Group Analogues (LGAs) in the local Universe (Sweet et al. 2026, in preparation). The overarching goal of the survey is to assess how the LG and its dwarf-galaxy population with total luminosities $M_V<-10$ compare with those of other galaxy groups, with particular emphasis on their spatially resolved star formation, chemical properties, and internal kinematics. As the survey is currently in its early phase, the primary objective of the current study is to establish, test, and validate robust data-reduction and analysis methodologies for dwarf galaxy observations obtained with KOALA\footnote{Kilofibre Optical AAT Lenslet Array, \url{https://aat.anu.edu.au/science/instruments/current/koala/overview}} and AAOmega\footnote{AAOmega spectrograph, \url{https://aat.anu.edu.au/science/instruments/current/AAOmega}} on the Anglo-Australian Telescope\footnote{Anglo-Australian Telescope, \url{https://aat.anu.edu.au/}} (AAT). In this context, the paper presents the first resolved spectroscopic results for three dwarf galaxies drawn from the LGA GAMA J1440--00 and the Choir group HIPASS J1159--19, while also contributing to the broader Delegate Survey effort through collaborative pipeline development and data processing. We also provide an initial comparison between the analysed LGA dwarf galaxies and LG dwarfs of similar luminosity.
\section{TARGET GALAXIES}

For this pilot study, which aims to test our data-processing methods and the sensitivity of the KOALA instrument, we selected the two dwarf irregular galaxies GAMA~79098 and 569709 from the GAMA J1440–00 Group, while the third galaxy HIPASS J1159–19 S2 (also known as NGC 4027A) was drawn from the gas-rich Choir Group\footnote{Galaxy groups comprising four or more emission line galaxies as identified by SINGG (Survey for Ionisation in Neutral Gas Galaxies; \citep{Meurer_2006}) imaging of HIPASS (H I Parkes All-Sky Survey; \citealp{HIPASS}) detections.} HIPASS J1159-19 (see Table~\ref{tab:Delegate}).

\begin{table*}

\begin{tabular}{|cccccccc|}
\hline
\multicolumn{1}{|c|}{Name} & \multicolumn{1}{c|}{GAMA/HIPASS} & \multicolumn{1}{c|}{RA}    & \multicolumn{1}{c|}{DEC}   & \multicolumn{1}{c|}{$m_T$} & \multicolumn{1}{c|}{$\mu_{D25}$ (B)} & \multicolumn{1}{c|}{$v_h$} & Reference                               \\
\multicolumn{1}{|l|}{}     & \multicolumn{1}{c|}{}            & \multicolumn{1}{c|}{(J2000)} & \multicolumn{1}{c|}{(J2000)} & \multicolumn{1}{c|}{(mag)}          & \multicolumn{1}{c|}{(mag\,arcsec$^{-2}$)}    & \multicolumn{1}{c|}{(\kms)}  &                                         \\
\multicolumn{1}{|c|}{(1)}  & \multicolumn{1}{c|}{(2)}         & \multicolumn{1}{c|}{(3)}   & \multicolumn{1}{c|}{(4)}   & \multicolumn{1}{c|}{(5)}          & \multicolumn{1}{c|}{(6)}                & \multicolumn{1}{c|}{(7)}   & (8)                                     \\ \hline
\multicolumn{8}{|c|}{GAMA J1440-00 Group (Central Galaxy Pair: NGC 5719/5713)}                                                                                                                                                                                      \\ \hline
NGC 5719                   & GAMA 64804                       & 14:40:56.4                & $-$00:19:05               & 11.38 (r)                         & 24.45                                   & $1733\pm3$                 & \cite{Barbara_2005}       \\
NGC 5713                   & GAMA 64771                       & 14:40:11.5               & $-$00:17:21              & 11.11 (r)                         & 22.22                                   & $1883\pm4$                 & \cite{deVacucoulurs_1991} \\ \hline
PGC 1159795                & GAMA~79098                       & 14:40:15.4                & 00:12:25               & 16.21 (r)                         & 23.59                                   & $1873\pm1$                 & \cite{DESI_2024}          \\
PGC 135857                 & GAMA 64758                       & 14:39:40.8                & $-$00:18:11              & 16.70 (r)                          & 25.56                                   & $1814\pm8$                  & \cite{DESI_2024}          \\
PGC 214313                 & GAMA 49173                       & 14:39:56.6                & $-$00:44:38              & 16.61 (r)                         & 24.21                                   & $1769\pm64$                & \cite{Colless_2001}       \\
PGC 1140314                & GAMA~569709                      & 14:40:15.4                & $-$00:33:43               & 17.62 (r)                         & ---                                     & $1766\pm15$                & \cite{DESI_2024}          \\
PGC 214314                 & GAMA 49204                       & 14:40:05.3               & $-$00:43:23               & 17.08 (r)                         & ---                                    & $1850\pm8$                 & \cite{DESI_2024}          \\
PGC 135858                 & GAMA 594420                      & 14:38:46.6               & $-$00:01:37               & 17.67 (r)                         & ---                                     & $1859\pm10$                & \cite{Roberts_2004}       \\
PGC 1155970                & GAMA 79084                       & 14:39:42.7                & 00:03:22                 & 17.70 (r)                         & ---                                     & $1722\pm8$                 & \cite{DESI_2024}          \\
AGC 540925                 & GAMA 594302                      & 14:38:46.6                & $-$00:01:37               & 17.66 (r)                         & ---                                     & $1832\pm8$                 & \cite{Haynes_2018}        \\
---                        & GAMA 64829                       & 14:40:54.2                & $-$00:21:22               & 19.47 (r)                         & ---                                     & $1764\pm5$                     & \cite{Baldry_2018}        \\ \hline
\multicolumn{8}{|c|}{HIPASS J1159-19 Group (Central Galaxy: NGC 4027)}                                                                                                                                                                                         \\ \hline
NGC 4027                   & HIPASS J1159-19                  & 11:59:30.2               &  $-$19:15:55               & 11.71 (B)                         & ---                                     & $1671\pm6$                 & \cite{deVacucoulurs_1991} \\ \hline
NGC 4027A                  & HIPASS~J1159-19~S2               & 11:59:29.3                & $-$19:20:00               & 15.03 (B)                         & ---                                     & $1700\pm55$                & \cite{deVacucoulurs_1991} \\
LEDA 856913                & HIPASS J1159-19 S3               & 11:59:36.0                & $-$19:19:03              & ---                               & ---                                     & ---                        & ---                                     \\
LEDA 856737                & HIPASS J1159-19 S4               & 11:59:37.9                & $-$19:19:46               & ---                               & ---                                     & ---                        & ---                                     \\ \hline
\end{tabular}

\caption{Galaxies in the galaxy groups GAMA J1440-00 and HIPASS J1159-19. (1) galaxy name; (2) GAMA or HIPASS ID; (3) right ascension (J2000); (4) declination (J2000); (5) apparent magnitude, with the photometric band given in brackets \citep{Lauberts_1989, Paturel_2003, 2MASS, Adelman-McCarthy_2008, Ahn_2012}; (6) mean $B$-band surface brightness within the $D_{25}$ isophote \citep{Moustakas_2023}; (7) heliocentric velocity; and (8) reference for velocity. Further details on galaxies in the GAMA~J1440--00 group are presented in \citet{Jerjen_2025}. 
}
\label{tab:Delegate}
\end{table*}

\begin{table}

    \centering

\begin{tabular}{cccll}
\hline
Target Name & RA & DEC & OET & SET \\ 
 & (J2000) & (J2000) &  (sec) & (sec) \\
\hline
GAMA~79098           & 14:40:15.2          & 00:12:23.6           & 18000            & 2400             \\
GAMA~569709          & 14:40:23.3          & $-$00:33:43.3         & 21600            & 3600             \\
HIPASS~J1159-19~S2 P1         & 11:59:30.0          & $-$19:20:09.2        & 18000            & 2400             \\
HIPASS~J1159-19~S2 P2          & 11:59:29.1          & $-$19:19:50.0        & 18000            & 2400             \\ \hline
\end{tabular}

    \caption{Chosen targets and the coordinates of their pointing centres. The object-frame exposure time (OET) and background sky exposure time (SET) for each target are listed.}
\label{tab:targets}
\end{table}

\begin{figure}

	\includegraphics[width=\columnwidth]{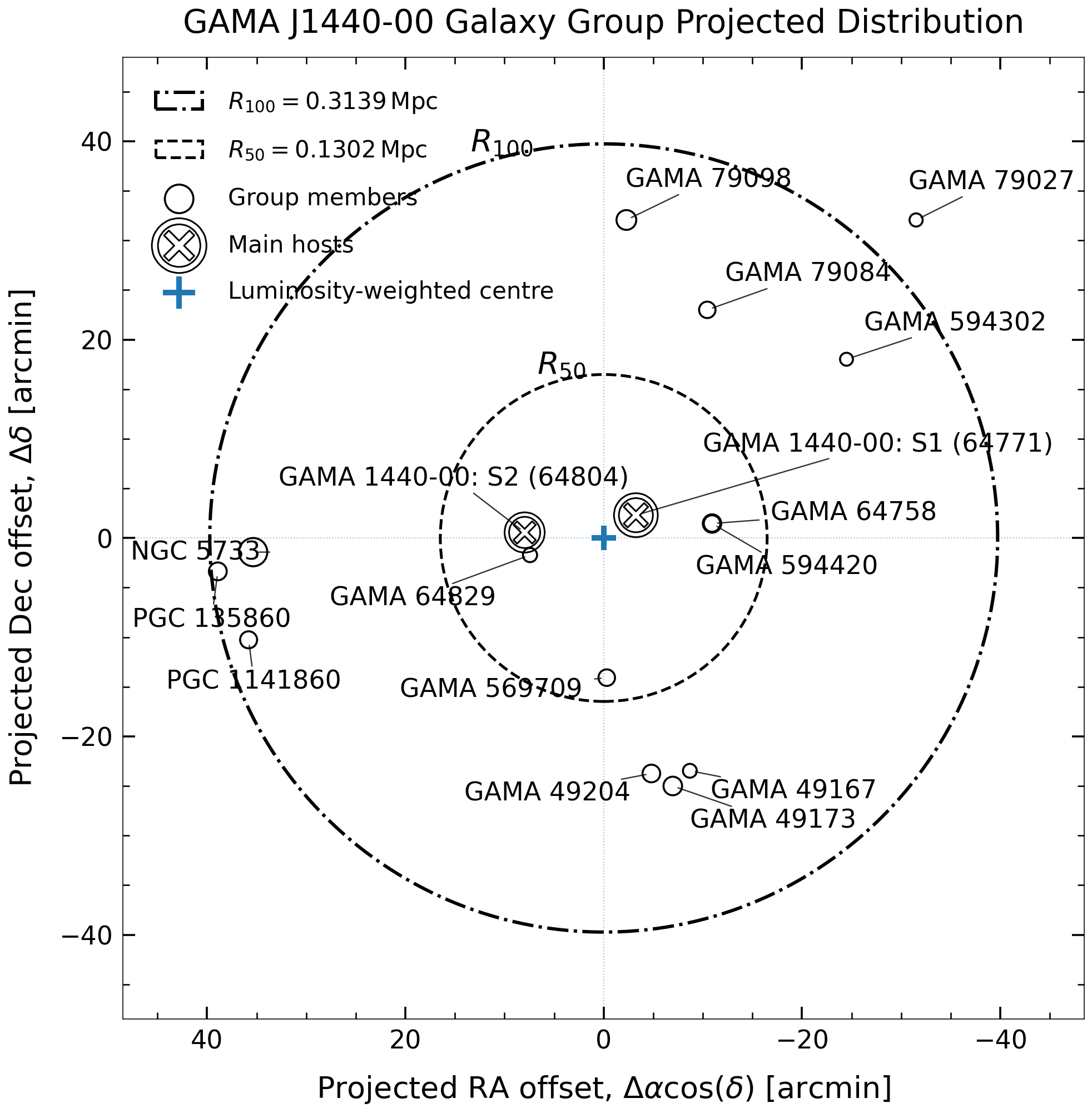}

    \caption{Projected distribution of the GAMA J1440-00 group dwarf galaxies targeted by the Delegate Survey. The $R_{100}$ and $R_{50}$ radii are defined by the projected distances of the outermost and median-ranked dwarf galaxies, respectively, from their nearest massive host, and are plotted as circles centred on the r-band luminosity-weighted centre of the system. The two massive spiral galaxies are indicated by large crossed circles, while the remaining dwarf members are shown as circles scaled according to their magnitudes. }
    \label{fig:GAMA_map}
\end{figure}

GAMA J1440–00 is a galaxy group hosting two interacting spiral galaxies, GAMA 64771 (NGC 5713) and GAMA 64804 (NGC 5719), along with 14 velocity-confirmed satellite galaxies \citep{Jerjen_2025}, 11 of which are targets of the Delegate Survey (see Fig. \ref{fig:GAMA_map}). This LGA is located at a distance of 27.16 Mpc \citep{Tulley_2016} and has an estimated virial radius of $r_{\mathrm{vir}} = 323^{+62}_{-52}$\,kpc \citep{GAMA_group, Jerjen_2025}.

GAMA~79098 has a heliocentric velocity of $1874 \pm 3$\,\kms \citep{DESI_2024} and a total absolute $B$-band magnitude of $-15.52$\,mag \citep{Colless_2001} when adopting the group distance of 27.16\,Mpc. The projected separation between GAMA~79098 and the group's nearest spiral galaxy NGC 5713 is 252.9\,kpc. 
GAMA~569709 has a velocity of $v_h=1766 \pm 15$\,\kms and a total absolute $B$-band magnitude of $-13.98$ \citep{DESI_2024, Colless_2001}. This galaxy has a projected distance of 127.8\,kpc from NGC 5713.

The Choir group HIPASS J1159–19 contains the prominent SB(s)dm spiral galaxy HIPASS J1159–19 S1 (NGC 4027) and three identified dwarf satellites, including HIPASS J1159–19 S2 (NGC 4027A), the bright dwarf galaxy located to the south of NGC 4027 \citep{Sweet_2013}. The distance to this group is 19.6 Mpc \citep{Pejcha_2015}. The observed velocity of the target is $v_h=1700 \pm 55$\,\kms and its total absolute $B$-band magnitude is $M_B=-16.43$ \citep{Lauberts_1989}. The projected physical separation from NGC 4027 is 28.06\,kpc.

\section{OBSERVATIONS \& DATA REDUCTION}
The KOALA instrument \citep{KOALA} is a wide-field integral-field unit that feeds the dual-beam, fibre-fed spectrograph AAOmega \citep{AAOmega} at the 3.9\,m AAT. KOALA employs a telecentric double-lenslet array comprising 1000 hexagonal microlenses, with interchangeable fore-optics providing selectable fields-of-view of $15.3 \times 28.3$\,arcsec$^{2}$ at 0.7\,arcsec sampling or $27.4 \times 50.6$\,arcsec$^{2}$ at 1.25\,arcsec sampling. The microlens-based fibre feed delivers a near-unity filling factor by eliminating fibre cladding losses, while improving fibre-coupling efficiency through pupil imaging at the fibre input.

Within AAOmega, the science fibres are reformatted into a pseudo-slit, collimated, and subsequently separated into blue and red arms using a dichroic beam splitter, selectable at 570 or 670\,nm. Each arm is equipped with an independent volume phase holographic transmission grating, with dispersed light recorded on $2048 \times 4096$ E2V CCD detectors, where spectral dispersion occurs along the $4k$ axis. AAOmega provides wavelength coverage spanning approximately 370–900\,nm and supports a suite of gratings delivering spectral resolving powers in the range $R \sim 1000$–$10,000$ \citep{Saunders_2004}.

For the target observations (completed in April-May 2024 and February 2025), the 570\,nm dichroic was employed in conjunction with the 580V and 1000R gratings on the blue and red arms, respectively. The 580V grating provides a resolving power of $R \simeq 1300$ and is centred at 4700\,\AA, delivering wavelength coverage from 3700 to 5700\,\AA. The 1000R grating provides a resolving power of $R \simeq 3400$ and is centred at 6880 and 9200\,\AA, yielding spectral coverage over the ranges 6280–7450\,\AA\ and 8650–9700\,\AA, respectively. This configuration enables simultaneous observations of prominent stellar absorption features, including the 4000\,\AA\ break (D4000), as well as strong nebular and auroral emission lines from ionised interstellar gas. These include [O\,\textsc{ii}] $\lambda\lambda$3727, 3729, H$\delta$, H$\gamma$, [O\,\textsc{iii}] $\lambda$4363, He\,\textsc{ii} $\lambda$4686, H$\beta$, [O\,\textsc{iii}] $\lambda\lambda$4959, 5007, [N\,\textsc{ii}] $\lambda$5755, [O\,\textsc{i}] $\lambda$6300, [S\,\textsc{iii}] $\lambda$6312, H$\alpha$, [N\,\textsc{ii}] $\lambda\lambda$6548, 6583, He\,\textsc{i} $\lambda$6678, [S\,\textsc{ii}] $\lambda\lambda$6717, 6731, [O\,\textsc{ii}] $\lambda\lambda$7318, 7330, and [S\,\textsc{iii}] $\lambda\lambda$9069, 9531.

To match the angular sizes of the target galaxies we used the larger field of view of $24\farcs7 \times 50\farcs6$. A total on-source exposure time of 18,000\,sec was allocated for each pointing, targeting a signal-to-noise ratio (SNR) of 10 per \AA\ in the blue continuum at a surface brightness level of 21.3\,B\,mag\,arcsec$^{-2}$. A small dithering pattern comprising five 1800\,s exposures for each red-arm configuration was employed to compensate for dead lenslets and mitigate the effects of cosmic rays. This resulted in five exposures for each red-arm setting and a total of 10 exposures in the blue arm. GAMA~79098 and 569709 each had a single KOALA pointing allocated, while HIPASS~J1159-19~S2 required two ($P1$ and $P2$). The centre coordinates for each pointing are given in Table~\ref{tab:targets}.\\

Five exposures were obtained with the spectrograph configured at central wavelengths of $\lambda = 4700$ \AA\,(blue arm) and $\lambda = 6880$ \AA\,(red arm), and a further five exposures at $\lambda = 4700$ \AA\,(blue arm) and $\lambda = 9200$ \AA\,(red arm). For each sequence of five object exposures, three 600\,s offset sky frames were interleaved between the dithered science frames, following the pattern $c$–sky–$a$–$b$–sky–$d$–$e$–sky, where $c$, $a$, $b$, $d$, and $e$ denote the individual dither positions. The present analysis is restricted to blue arm and red arm observations centred at $\lambda = 6880$ \AA.

The initial data reduction was performed using the \textsc{2dfdr}\footnote{\url{aat.anu.edu.au/science/software/2dfdr}} software package. This pipeline provides automated calibration and reduction of raw CCD frames using instrument-specific configuration files, executing the standard processing steps required for fibre spectroscopy. These include bias subtraction, dark-current correction, CCD flat-fielding, fibre trace (tramline) identification, optimal spectral extraction, wavelength calibration using arc-lamp exposures, fibre throughput and spatial flat-field corrections, sky subtraction, and the combination of repeated calibration exposures \citep{2dfdr}. All observations were reduced independently for the blue and red arms and for each observing night. Following reduction, the extracted spectra are output in the form of Row-Stacked Spectra (RSS) files, which serve as input for \textsc{PyKOALA}. 

\textsc{PyKOALA} \citep{Pykoala} is used to convert RSS files into fully calibrated three-dimensional data cubes, providing wavelength, flux, and astrometric solutions suitable for scientific analysis. We adopt a \textsc{PyKOALA}-based reduction workflow developed by Ángel R. López-Sánchez, which extends the core \textsc{PyKOALA} functionality with optimised calibration procedures and quality-control diagnostics tailored to KOALA observations. 

Nightly calibration products are derived independently for each spectrograph arm and wavelength configuration. Fibre throughput variations are characterised using sky-flat RSS frames, from which a two-dimensional throughput map is extracted and used for correction.
Minor residual wavelength offsets ($\sim 0.1$\, \AA) between fibres are corrected using bright night-sky emission lines measured in sky RSS frames, refining the initial arc-based wavelength solution produced by \textsc{2dfdr}. Telluric absorption correction is derived from spectrophotometric standard star observations and applied to the red arm data, where atmospheric absorption features are present. Flux calibration is performed using standard stars by comparing observed counts to intrinsic stellar spectra to derive a wavelength-dependent instrumental response function. 

Science and associated sky RSS frames are calibrated using the derived nightly calibrations. As sky brightness varies with time and differs between continuum and emission-line components, sky subtraction is performed using independently optimised scaling factors for the sky continuum and sky emission lines, determined iteratively for each science–sky exposure pair. For the blue-arm data, where no strong sky emission lines are present, only continuum scaling is applied. 

The final data cubes are constructed by reprojecting the RSS data into the spatial image plane and combining individual exposures. Atmospheric differential refraction (ADR) and small spatial offsets between dithered exposures are measured during cube reconstruction using either a field star or the galaxy centre as a reference. Although the measured ADR is typically small compared to the spatial pixel scale, ADR corrections are applied when the wavelength-dependent offsets become comparable to or exceed the pixel size. The relative alignment between blue and red cubes is verified using common spatial features, and small inter-arm shifts are corrected when necessary to ensure consistent spatial registration.
\begin{figure}

	\includegraphics[width=\columnwidth]{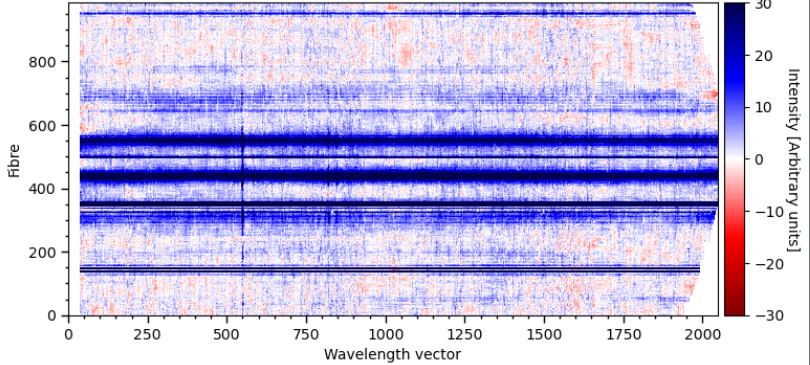}
    \includegraphics[width=0.9\columnwidth]{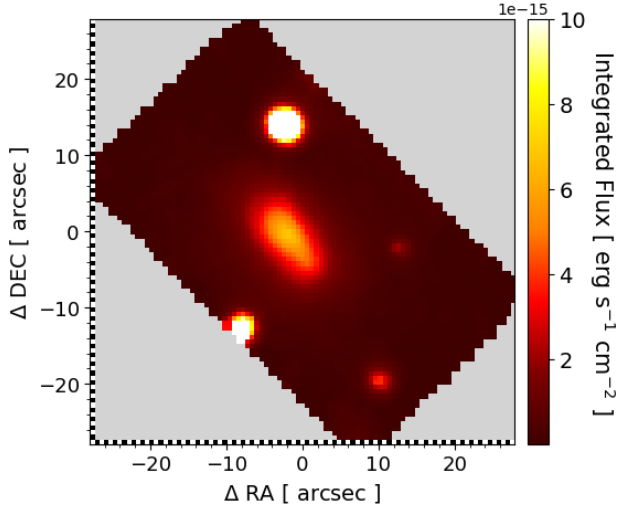}
    \includegraphics[width=0.9\columnwidth]{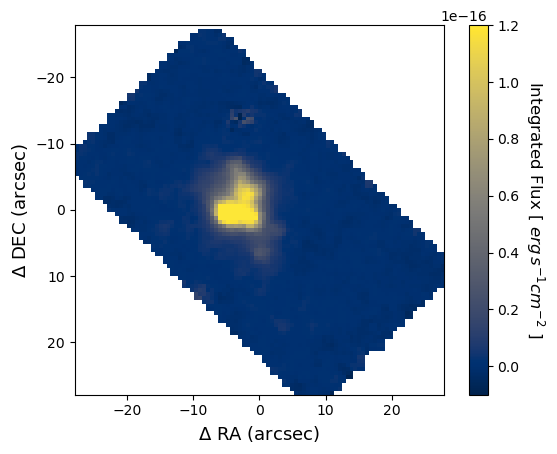}
    \caption{Figures illustrating the results of the data reduction and cube reconstruction for GAMA~79098. Top: Row-stacked spectra (RSS) from a single object exposure after full calibration and sky subtraction. Middle: Two-dimensional image of the total integrated flux across all spaxels in the reconstructed data cube, obtained from the red-arm observations. Bottom: Continuum-subtracted emission-line map for [O\,\textsc{iii}] $\lambda\lambda4959, 5007$\,\AA. }
    \label{fig:observation}
\end{figure}

\section{METHODS AND DATA ANALYSIS}

This section describes the analysis techniques used to derive the physical and kinematic properties of the target galaxies from the reduced integral-field spectroscopy data. We outline the procedures adopted for spectrum extraction, extinction correction, stellar continuum and emission-line fitting, and the derivation of star formation rates, gas-phase metallicities, and stellar population properties. We further describe the methods used to construct two-dimensional velocity fields and to quantify galaxy kinematics using harmonic modelling. All measurements are based on spatially resolved data and are designed to ensure robustness against low SNR regions and systematic uncertainties.

\subsection{Spectrum Extraction}
To obtain galaxy spectra suitable for measuring gas-phase metallicities, star formation rates, luminosity-weighted stellar ages, and mass-to-light ratios, we co-added all spaxels associated with each galaxy. During this process, spaxels that did not contribute to an improvement in the overall SNR when combined were excluded, along with those associated with obvious foreground sources. To identify spaxels with sufficient SNR, we first evaluated the median continuum SNR of each spaxel and masked those below a specified threshold. The spectra of the remaining spaxels were then co-added to determine the combined SNR. The threshold was progressively lowered and the combined SNR re-evaluated until the inclusion of additional spaxels caused the combined SNR to decrease.

For galaxies GAMA~79098 and HIPASS J1159–19 S2 we also identified spaxels corresponding to regions of intense star-forming activity. Spaxels dominated by star-formation-driven ionisation were identified using the Baldwin–Phillips–Terlevich (BPT; \citealt{Baldwin_1981}) diagnostic applied on a spaxel-by-spaxel basis in regions with significant H$\alpha$ emission (SNR > 3). Spectra were extracted separately for these regions, in addition to the galaxy-wide co-added spectra. 
\begin{figure*}

  \includegraphics[width=0.28\linewidth, height = 0.27\linewidth]{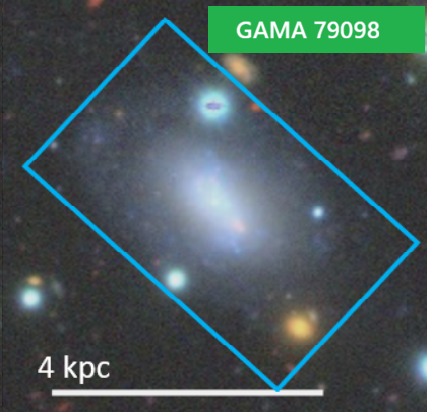}
  \includegraphics[width=0.28\linewidth, height=0.27\linewidth]{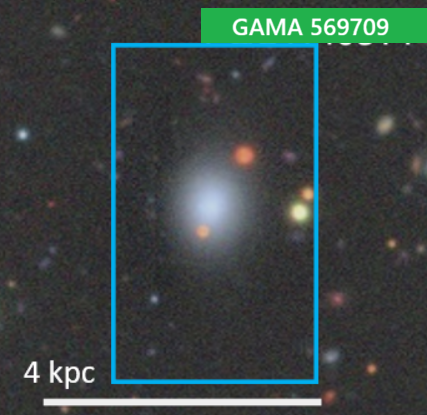}
  \includegraphics[width=0.28\linewidth, height=0.27\linewidth]{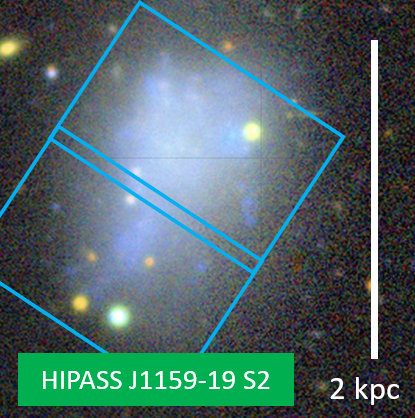}
  \\
  
  \includegraphics[width=0.3\linewidth, height=0.29\linewidth]{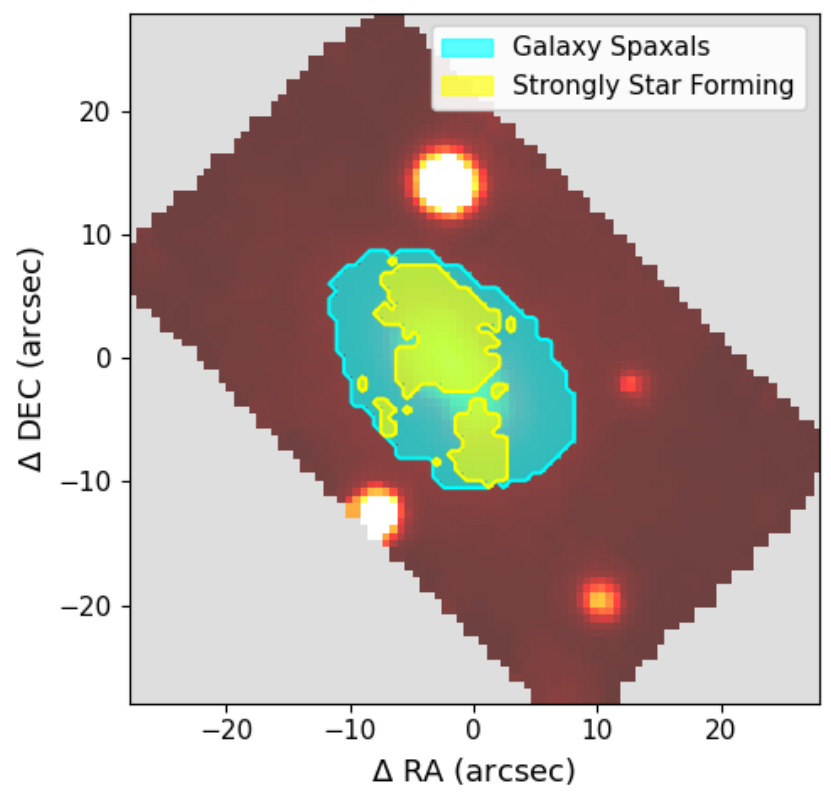}
  \includegraphics[width=0.28\linewidth, height=0.29\linewidth]{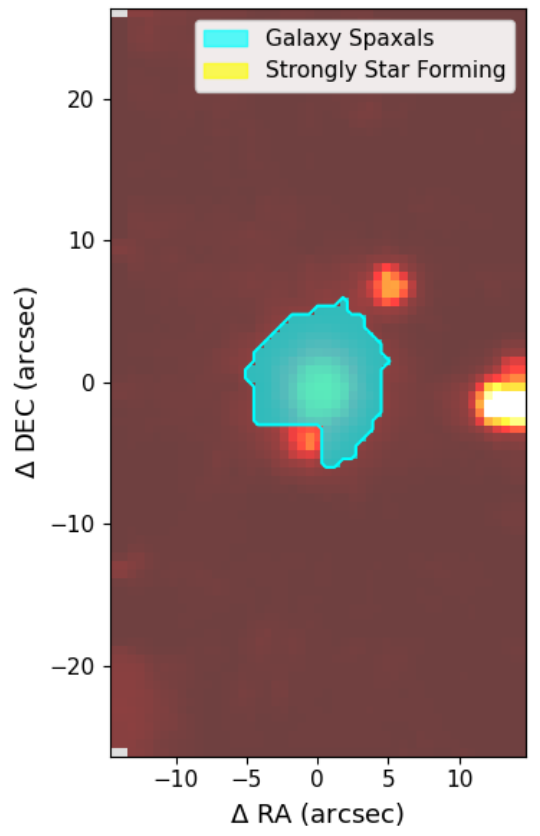}
  \includegraphics[width=0.3\linewidth, height=0.29\linewidth]{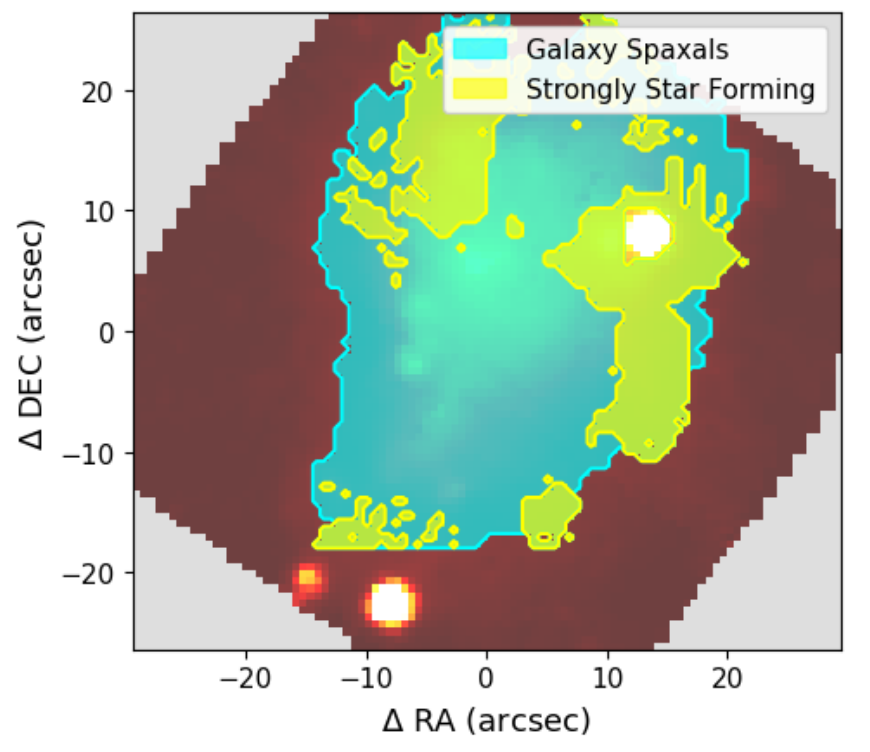}
  \includegraphics[width=0.05\linewidth, height=0.29\linewidth]{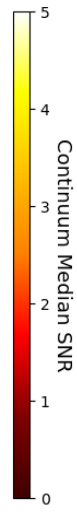}
  \\

  \caption{\textbf{Top row:} HSC–SSP DR2 composite images of GAMA~79098 (left) and GAMA~569709 (middle), and a Legacy Survey DR10 composite image of HIPASS J1159–19 S2 (right). The horizontal scale bar corresponds to approximately 4\,kpc, while the vertical scale bar corresponds to 2\,kpc. Blue, green, and red colours represent the observed $g$, $r$, and $z$ bands, respectively. Blue rectangular outlines indicate the KOALA field of view for each exposure. \textbf{Bottom row:} Spaxels used for extracting the integrated galaxy spectra (highlighted blue), overlaid on maps of the median continuum SNR, for GAMA~79098 (left; 747 spaxels), GAMA~569709 (middle; 208 spaxels), and HIPASS J1159–19 S2 (right; 3297 spaxels). Spaxels with insufficient spectral SNR or contaminated by foreground sources are excluded. Spaxels classified as strongly star-forming based on the [N II]-BPT diagnostic and H$\alpha$ emission are highlighted in yellow. }
  \label{fig:targets}
\end{figure*}

\subsection{Extinction Correction}

The foreground extinction caused by dust within the MW is corrected using $E(B-V)$ values obtained from the \textit{Galactic Dust Reddening and Extinction Service}\footnote{\url{https://irsa.ipac.caltech.edu/applications/DUST/docs/background.html}}. We apply the MW extinction model presented by \citet{Gordon_2023}, implemented via the publicly available Python \texttt{dust\_extinction} package \citep{Gordon_2024}. The extinction curve is parameterised using the adopted $E(B-V)$ values and 
$R_V = A_V / E(B-V) = 3.1$, appropriate for the MW \citep{Gordon_2023}.

Internal dust extinction was estimated for GAMA~79098 and HIPASS J1159–19 S2 using the H$\beta$ / H$\gamma$ Balmer decrement \citep{Dopita_2003_book}. In contrast, GAMA~569709 exhibits no detectable Balmer emission or other nebular emission lines sufficient for determining a Balmer decrement (see Fig. \ref{fig:spectra}), consistent with a passive system, and therefore no internal extinction correction was applied. For GAMA~79098 and HIPASS J1159–19 S2, extinction corrections were performed using the Small Magellanic Cloud (SMC) extinction model, adopting the \textit{G24 SMCAvg} curve \citep{Gordon_2024b}, as both target galaxies exhibit luminosities and gas-phase metallicities comparable to those of the SMC.

\subsection{Spectral Fitting}
\label{sec:spectral fitting}
To analyse the co-added blue-arm spectra, we employ the penalised pixel-fitting (\textsc{pPXF}) algorithm, a full-spectrum fitting method implemented in Python that uses a penalised maximum-likelihood approach to fit stellar population synthesis (SPS) models to galaxy spectra \citep{Cappellari_2004, Cappellari_2017, Cappellari_2023}. For spectra with sufficient SNR ($>20$ for continuum), pPXF enables the extraction of detailed stellar population properties, including age and metallicity distributions and stellar mass-to-light ratios. The SPS templates adopted in this work are based on version~3.2 of the Flexible Stellar Population Synthesis (FSPS) models\footnote{\url{https://github.com/cconroy20/fsps}}
, calibrated following \citet{Cappellari_2023}. These models span stellar ages as young as 1\,Myr, making them well suited to fitting galaxies in our sample that are actively forming stars \citep{fsps2,fsps1}.

Our \textsc{pPXF} fitting procedure closely follows the most recent guidelines provided by \cite{Cappellari_2023}. In particular, we adopt a two-step “fit-and-clean” approach to improve fit robustness for spectra with relatively low continuum SNR, with modest regularisation. In the first step, an initial wavelength mask is applied to exclude narrow regions around strong emission lines and noisy spectral edges, and an initial \textsc{pPXF} fit is performed. The residuals from this fit are then compared to the observational uncertainties, derived from the flux variance, to identify outlier pixels where the residual exceeds three times the estimated error. These outlier regions are incorporated into an updated mask, and a final \textsc{pPXF} fit is performed. This procedure reduces the sensitivity to extreme outliers and noisy spectral regions, yielding more stable results for spectra with low SNR.

From the resulting SPS template weights, we infer luminosity- and mass-weighted distributions of stellar age and metallicity for each galaxy. These distributions are used to reconstruct the star formation history and to estimate the mean stellar metallicity. In addition, the best-fitting model spectrum and associated template weights are used to derive stellar mass-to-light ratios and stellar-only luminosities in the SDSS $r$ and Johnson $V$ bands.

Uncertainties on fitted parameters are estimated using a wild bootstrap resampling technique with residual perturbation, following \citet{Davidson_2008} and the recommendations of \citet{Cappellari_2023}. In this approach, the fit residuals are randomly multiplied by $+1$ or $-1$ with equal probability before resampling, rather than being directly resampled as in standard residual bootstrapping. The uncertainties are estimated as the standard deviation of the fitted parameters obtained from multiple bootstrap realisations.

For the red-arm spectra, the stellar continuum is estimated using an 11th-order polynomial fit, which yields the best performance among polynomial orders in the range 3-20, as the \textsc{pPXF} fitting proved unstable or produced poor results. This behaviour is attributed to the lack of strong stellar absorption features in the red-arm spectra of the target dwarf galaxies, combined with residual sky features at a level comparable to weak absorption lines. As in the blue arm, prominent emission lines and spectral edges are masked prior to fitting. The polynomial continuum fits in the red arm are found to be stable owing to the smoother continuum and lower noise levels, whereas in the blue arm the presence of strong absorption features necessitates the use of \textsc{pPXF} where feasible.

\begin{figure*}

  \includegraphics[width=\linewidth, height = 0.22\linewidth]{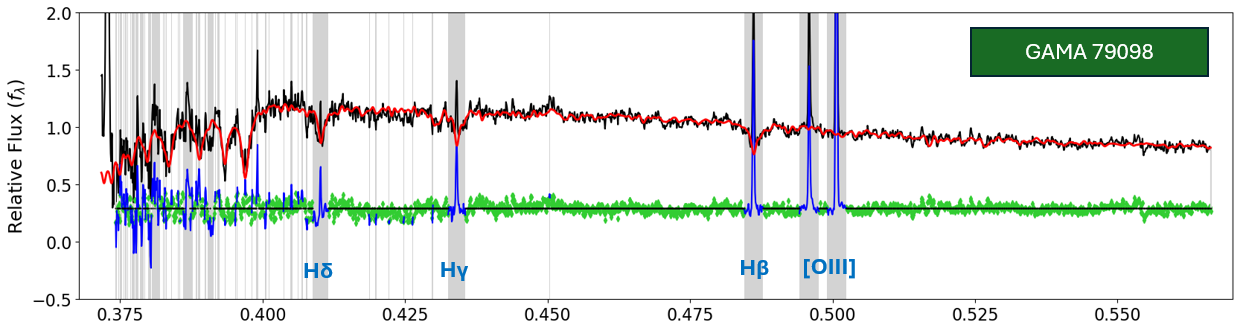}
  \includegraphics[width=\linewidth, height=0.22\linewidth]{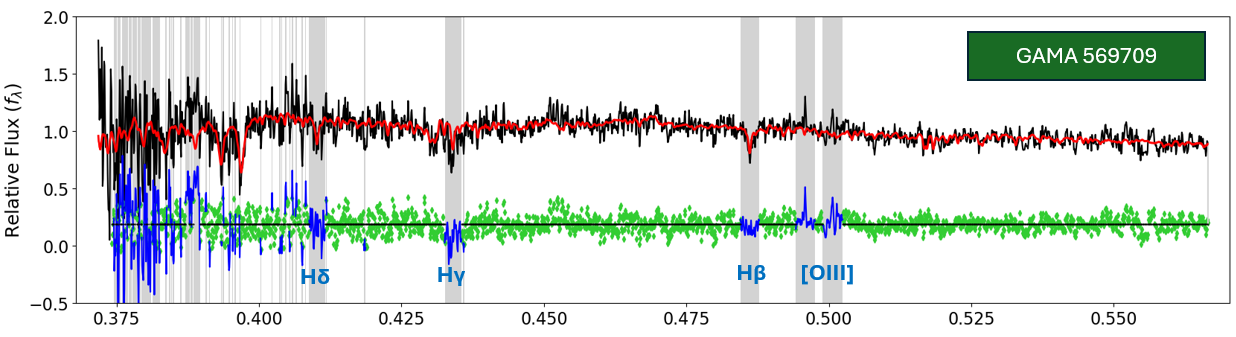}
  \includegraphics[width=\linewidth, height=0.24\linewidth]{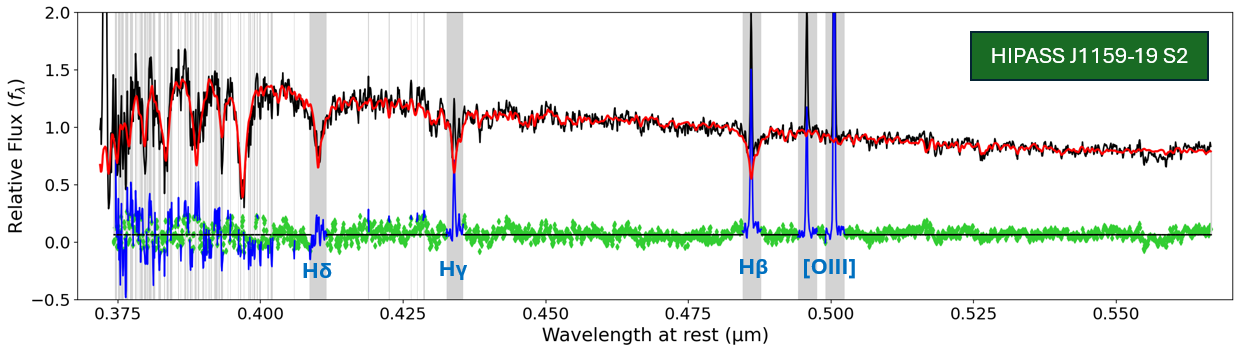}

  \caption{Model spectra from the best-fitting \textsc{pPXF} results (red curves) overlaid on the observed galaxy spectra (black curves) for GAMA~79098 (top), GAMA~569709 (middle), and HIPASS J1159–19 S2 (bottom). The spaxels used to construct the combined galaxy spectra are shown in Fig.~\ref{fig:targets}. Grey vertical bands indicate wavelength regions masked prior to the final \textsc{pPXF} fit, corresponding to prominent gas-phase emission lines or spectral regions where the initial fit quality was poor (outlier pixels). The masked emission lines include H$\beta$, H$\gamma$, H$\delta$, and [O\,\textsc{iii}] $\lambda\lambda4959, 5007$\,\AA. The green points and blue spectrum show the fit residuals, which are used to identify extreme outliers during the ‘fit-and-clean’ procedure. The blue residual spectrum is subsequently used to measure the gas-phase emission-line properties.
  }
    \label{fig:spectra}
\end{figure*}

The results of the \textsc{pPXF} fits for the three analysed dwarf galaxies are presented in Fig.~\ref{fig:spectra}. For HIPASS J1159–19 S2, the best-fitting stellar model reproduces the main absorption features, including the Ca\textsc{ii} K and H lines and the Mg\,\textit{b} triplet, enabling a robust determination of the stellar metallicity. In contrast, the lower continuum SNR of the GAMA group targets leads to less reliable recovery of absorption features and correspondingly larger uncertainties (i.e. approximately 20 times larger uncertainty) in the inferred stellar metallicities.

\subsection{Emission Line Study}

Gas-phase emission-line fluxes are measured from the continuum-subtracted spectra. Our standard approach for flux measurements is direct numerical integration over a fixed wavelength interval centred on each emission line, defined as four times the full width at half-maximum (FWHM) determined from a Gaussian fit to the line profile. For per-spaxel emission-line measurements used in the construction of two-dimensional BPT diagnostics, the blue-arm continuum is estimated using an order 11 polynomial fit, as the spectra of individual spaxels generally have insufficient SNR for reliable \textsc{pPXF} fitting.

Gas-phase metallicities are estimated from strong emission-line ratios measured from co-added spectra extracted exclusively from spaxels classified as star-forming regions. For reliable measurements of oxygen abundance, expressed as $12 + \log(\mathrm{O/H})$, in the range $7.0$--$8.5$, we adopt the calibration of \citet{Pilyugin_2016}, which uses a combination of sulphur, oxygen, nitrogen and hydrogen emission lines. The line-ratio definitions are given in Eqn.~\eqref{eq:line_ratios}, with the upper- and lower-branch formulations of the $S$ calibration presented in Eqns.~\eqref{eq:S_U} and~\eqref{eq:S_L}, respectively. Regions with $\log(N_2) > -0.6$ are assigned to the upper branch, while those with $\log(N_2) \leq -0.6$ are assigned to the lower branch.


\begin{equation}
\label{eq:line_ratios}
\begin{aligned}
R_2 &= \frac{I_{[\mathrm{O\,II}]\lambda3727+3729}}{I_{\mathrm{H}\beta}}, \quad
R_3 &&= \frac{I_{[\mathrm{O\,III}]\lambda4959+5007}}{I_{\mathrm{H}\beta}}, \\
S_2 &= \frac{I_{[\mathrm{S\,II}]\lambda6717+6731}}{I_{\mathrm{H}\beta}}, \quad
N_2 &&= \frac{I_{[\mathrm{N\,II}]\lambda6548+6583}}{I_{\mathrm{H}\beta}}
\end{aligned}
\end{equation}

\begin{equation}
\label{eq:S_U}
\begin{aligned}
&12 + \log(\mathrm{O}/\mathrm{H})_{S,U}
 = 8.424 
  + 0.030\,\log\bigl(\tfrac{R_{3}}{S_{2}})
  + 0.751\,\log N_{2} \\
&\quad  
+ (-0.349
  + 0.182\,\log(\tfrac{R_{3}}{S_{2}})
  + 0.508\,\log N_{2}\Bigr)\,\log S_{2}
\end{aligned}
\end{equation}

\begin{equation}
\label{eq:S_L}
\begin{aligned}
&12 + \log(\mathrm{O}/\mathrm{H})_{S,L}
= 8.072
  + 0.789\,\log(\tfrac{R_{3}}{S_{2}})
  + 0.726\,\log N_{2} \\
&\quad
  + (1.069
  - 0.170\,\log(\tfrac{R_{3}}{S_{2}})
  + 0.022\,\log N_{2})\,\log S_{2}
\end{aligned}
\end{equation}

The star formation rates (SFRs) of the target galaxies are inferred from their H$\alpha$ luminosities. We estimate SFRs using the calibration relation provided by \citet{Calzetti_2007}, given in their equation~6:
\begin{equation}
\label{eq:SFR}
\mathrm{SFR},(M_\odot\,\mathrm{yr}^{-1}) = 5.3 \times 10^{-42} L(\mathrm{H}\alpha){\mathrm{corr}} (\mathrm{erg\,s}^{-1}),
\end{equation}
where $L(\mathrm{H}\alpha){\mathrm{corr}}$ is the H$\alpha$ luminosity corrected for dust extinction. This relation is an updated version of the calibration presented by \citet{Kennicutt_1998}, incorporating the 2005 update to the Starburst99 stellar population models \citep{Leitherer_1999}.

\subsection{Velocity Fields}

Line-of-sight (LOS) velocities are determined by fitting a Gaussian profile to the H$\alpha$ emission line in spaxels with sufficient H$\alpha$ SNR ($\geq3$). For two-dimensional velocity mapping, a pixel-by-pixel Gaussian fitting procedure is applied, and the derived velocity peaks are referenced to the mean velocity obtained from luminosity-weighted peak position measured from the co-added spectrum.

To obtain high-resolution velocity measurements at sub-pixel scales, we implement a peak-finding workflow optimised for low-SNR data. To suppress high-frequency noise, a Savitzky–Golay filter \citep{SG_1964} is applied to a narrow wavelength window around the estimated line centre using the SciPy \textit{savgol filter} module\footnote{\url{https://docs.scipy.org/doc/scipy/reference/generated/scipy.signal.savgol_filter.html}}
. The smoothed spectrum is then interpolated using a cubic spline, implemented through the SciPy \textit{CubicSpline} module\footnote{\url{https://docs.scipy.org/doc/scipy/reference/generated/scipy.interpolate.CubicSpline.html}}
, allowing for a refined estimate of the local line maximum \citep{Unser_1999}. Subsequently, a Gaussian profile is fitted using this interpolated peak as the initial centroid.

LOS velocity shear ($V_S$) is defined as half the difference between the 95th percentile ($v_{\mathrm{P95}}$) and the 5th percentile ($v_{\mathrm{P05}}$) of the velocity distribution (see Eqn~\eqref{eq:V_S1}), providing a robust estimate of the velocity amplitude that is insensitive to extreme outliers. To characterise the velocity dispersion ($\sigma_V$) while further minimising the influence of outliers, we compute a robust dispersion based on the median absolute deviation (MAD). This dispersion, denoted $\sigma_{\mathrm{MAD}}$, is calculated from the per-spaxel velocities ($v_i$) relative to the median velocity ($v_{\mathrm{med}}$), as defined in Eqn.~\eqref{eq:V2}. The uncertainty in the velocity measurements arising from the peak-finding procedure is propagated accordingly to assess the intrinsic significance of the measured velocity dispersion. A scaling factor of $k_c = 1.4826$ is adopted under the assumption of an approximately normal velocity distribution \citep{MAD}.

\begin{eqnarray}
\label{eq:V_S1}
   & V_S = (v_{(P95)}-v_{(P05)})/2 \\
   \label{eq:V2}
   &\sigma_{MAD} = k_c \times median(|v_i-v_{med}|)
\end{eqnarray}

To quantify the rotational properties of the galaxy line-of-sight velocity distribution (LOSVD), we perform the harmonic ring model fitting following \citet{Krajnovic_2006}. The LOSVD is decomposed into a set of concentric elliptical rings, with the velocity field on each elliptical ring modelled as a finite series of harmonic terms (see equation~\ref{eq:vharm}). For a given ring, the velocity at the semi-major axis radius $a$ and the eccentric anomaly $\psi$ are described by the harmonic coefficients $A_n$ and $B_n$, where $n$ denotes the harmonic order. The kinematic properties of a galaxy, including its rotation centre, velocity, PA, and inclination, are best constrained by minimising the harmonic coefficients.

The optimal representation of the velocity field is obtained by fitting the first few harmonic terms while optimising the associated geometric parameters. Previous studies have shown that harmonic terms of order $n>3$ are largely insensitive to inaccuracies in the ellipse parameters, such as the adopted centre, flattening, or position angle \citep{Krajnovic_2006}. We verified this behaviour for our sample by explicitly including higher-order ($n>3$) terms and found that they do not lead to a statistically significant improvement in the fits. Consequently, we adopt a harmonic expansion truncated at order $n=3$ to characterise the rotational structure of the LOSVD. 

The rotational amplitude ($V_{{rot}}$) and kinematic position angle ($PA$) for each ring are derived from the first-order harmonic coefficients, as defined in equations~\eqref{eq:gal_v} and~\eqref{eq:gal_PA}. The rotational properties derived from the harmonic model account for the inclination of the velocity field \citep{Krajnovic_2006} and are therefore compared with inclination-corrected rotational velocities of LG dwarfs in the comparison sample.

\begin{eqnarray}
\label{eq:vharm}
    V(a\,\psi) = A_0(a)+\Sigma_{n=1}^N A_n(a)\sin(n\psi)+B_n(a)\cos(n\psi)\\
\label{eq:gal_v}
    V_{rot} = k_1 = \sqrt{A_1^2+B_1^2}\\
\label{eq:gal_PA}
    PA_{gal} = \phi_1 = \arctan(A_1/B_1)
\end{eqnarray}

The harmonic models provide a satisfactory representation of the observed velocity fields but become less reliable towards the outer regions of the LOSVD, where the limited number of valid spaxels sampling each elliptical ring leads to increased uncertainty in the fits. Rings located near the edges of the velocity field that yield rotation velocities significantly exceeding the global velocity shear are therefore considered unreliable and are excluded from further analysis. 

To estimate the uncertainties associated with the kinematic measurements, we adopt a Monte Carlo (MC) resampling approach. Analogously to the bootstrap procedure used for the \textsc{pPXF} fitting, the velocity maps are resampled by perturbing each spaxel velocity according to its associated measurement uncertainty derived from Gaussian peak fitting. The kinematic fitting procedure is repeated for each resampled realisation, and the uncertainties on the derived parameters are taken as the standard deviations of their distributions across multiple MC trials.

\section{RESULTS}

This section presents the results derived from the analysis of the reduced integral-field spectroscopy data for the target dwarf galaxies. We report measurements of stellar population properties, including stellar masses, star formation histories, and stellar metallicities, alongside gas-phase characteristics such as internal extinction, star formation rates, and gas-phase metallicities. We also present spatially resolved kinematic measurements, including LOSVD and rotational properties inferred from harmonic modelling of the velocity maps. The quoted uncertainties are derived from the propagated variance in the reduced data cubes and from resampling-based techniques, including bootstrap and Monte Carlo methods (see Sec~\ref{sec:spectral fitting}).

\subsection{Stellar Properties}
\subsubsection{GAMA~79098}
The stellar-only absolute magnitude of GAMA~79098 in the SDSS $r$ band, as inferred from the SPS modelling, is $M_r = -16.5 \pm 0.4$. The corresponding calibrated $r$-band stellar mass-to-light ratio in solar units is $(M/L)_r = 1.2 \pm 0.3~(M_\odot/L_{\odot,r})$, implying a total stellar mass of $\log(M_{\star}/M_\odot) = 8.5 \pm 0.2$. This value is slightly higher than the stellar mass obtained from the GAMA photometric SED-fitting analysis, which yields $\log(M_{\mathrm{gal}}/M_\odot) = 8.2 \pm 0.12$\,dex \citep{GAMA_stellarmass}.

The star formation history (SFH) of GAMA~79098 is characterised by two relatively distinct star-forming episodes (see Fig.~\ref{fig:stellar}), the first occurring during the early stages of galaxy assembly, approximately 10–14 Gyr ago, and contributing about 40\% of the total stellar mass. This is followed by multiple star-forming peaks, culminating in the current episode of strong ongoing star formation, with stellar ages $\leq$ 6 Gyr, accounting for approximately 60\% of the total stellar mass. The presence of active star formation is further supported by the detection of strongly ionised H\,\textsc{ii} regions exhibiting prominent H$\alpha$ emission.

GAMA~79098 hosts a composite stellar population spanning a wide range of metallicities (see Fig.~\ref{fig:stellar}), from metal-poor ([Fe/H] $\leq -1.75$) to approximately solar metallicity ([Fe/H] $\simeq 0$). The older stellar population (ages $>10$\,Gyr), associated with the initial starburst, is dominated by metal-poor stars, whereas the younger population formed during the ongoing star formation episode includes both metal-rich and metal-poor components. The mean stellar metallicity on the solar scale is $\langle\mathrm{[Fe/H]}\rangle = -0.7 \pm 0.3$\,dex.

\subsubsection{GAMA~569709}
GAMA~569709 exhibits a stellar-only $r$-band absolute magnitude of $M_r = -14.4 \pm 0.5$ derived from the spectroscopic SPS fitting, with a corresponding stellar mass-to-light ratio of $(M/L)_r = 2.2 \pm 0.3~(M_\odot/L_{\odot,r})$. These values imply a total stellar mass of $\log(M_\star/M_\odot) = 8.0 \pm 0.3$, in good agreement with the stellar mass derived from photometric SED fitting, $\log(M_\star/M_\odot) = 8.0 \pm 0.13$ \citep{GAMA_stellarmass}.

The SFH of GAMA~569709 is dominated by an intense phase of star formation during the early stages of its assembly  (see Fig.~\ref{fig:stellar}), approximately 10–14\,Gyr ago, followed by a more recent starburst episode. Its stellar metallicity distribution is broadly similar to that of GAMA~79098, comprising both metal-poor and metal-rich stellar populations. The older stellar component associated with the initial starburst is dominated by metal-poor stars, whereas the younger population formed more recently is comparatively metal-rich. The mean stellar metallicity is $\langle\mathrm{[Fe/H]}\rangle = -0.7 \pm 0.4$\,dex.

\subsubsection{HIPASS~J1159-19~S2}
The stellar-only $r$-band absolute magnitude of HIPASS J1159–19 S2 is measured to be $M_r = -17.9 \pm 0.6$. Photometry-based SED-fitting results for this galaxy are not available in the literature. The derived calibrated $r$-band stellar mass-to-light ratio is $0.15 \pm 0.04~(M/L)_{\odot,r}$, yielding a total stellar mass of $\log(M_\star/M_\odot) = 7.8 \pm 0.3$.

\begin{figure*}

  \includegraphics[width=0.33\linewidth, height = 0.25\linewidth]{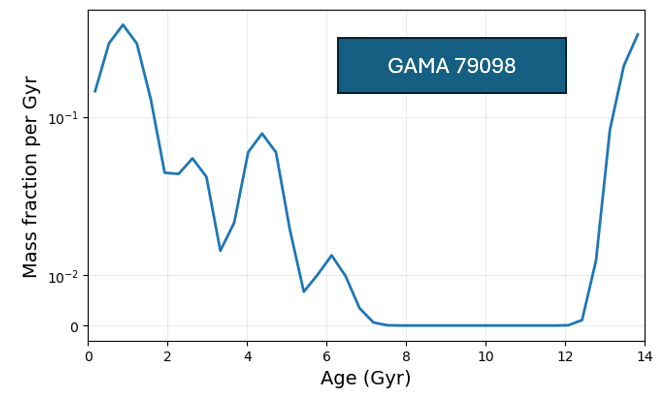}
  \includegraphics[width=0.33\linewidth, height=0.25\linewidth]{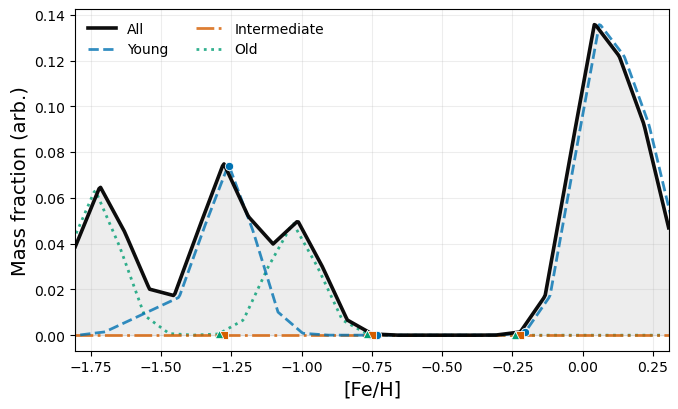}
  \includegraphics[width=0.33\linewidth, height=0.25\linewidth]{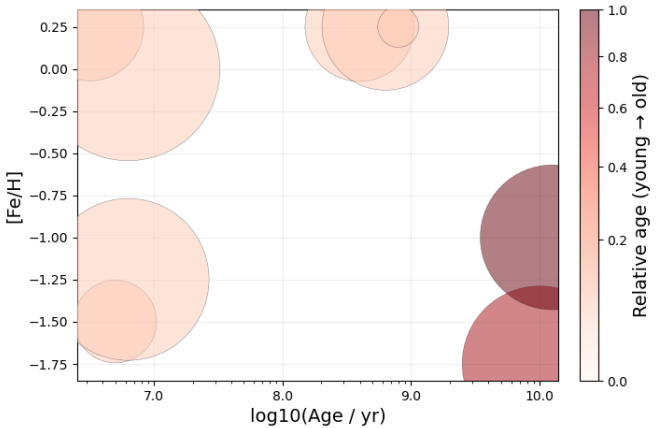}

  \includegraphics[width=0.33\linewidth, height = 0.25\linewidth]{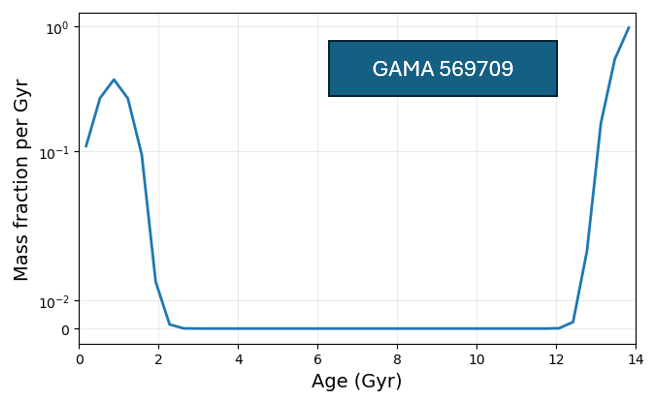}
  \includegraphics[width=0.33\linewidth, height=0.25\linewidth]{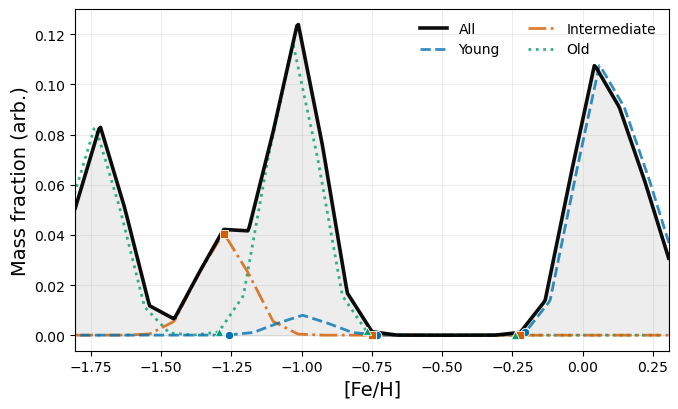}
  \includegraphics[width=0.33\linewidth, height=0.25\linewidth]{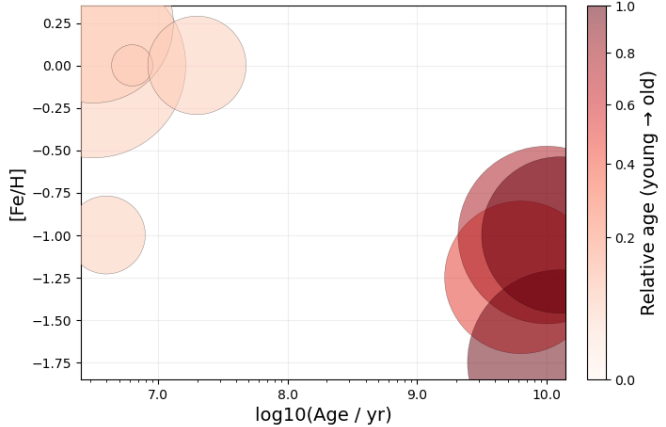}

  \includegraphics[width=0.33\linewidth, height = 0.25\linewidth]{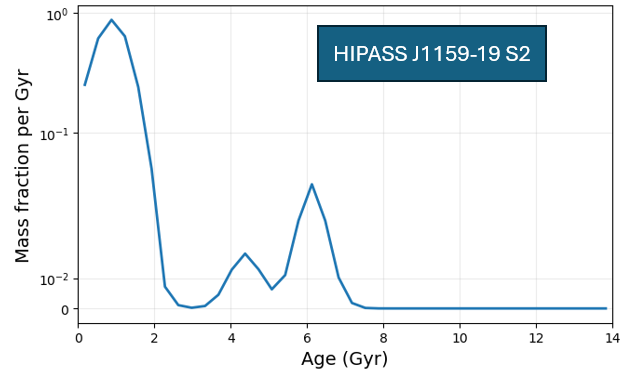}
  \includegraphics[width=0.33\linewidth, height=0.25\linewidth]{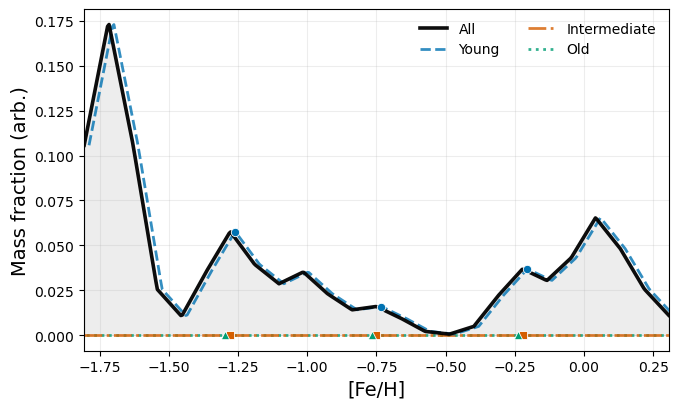}
  \includegraphics[width=0.33\linewidth, height=0.25\linewidth]{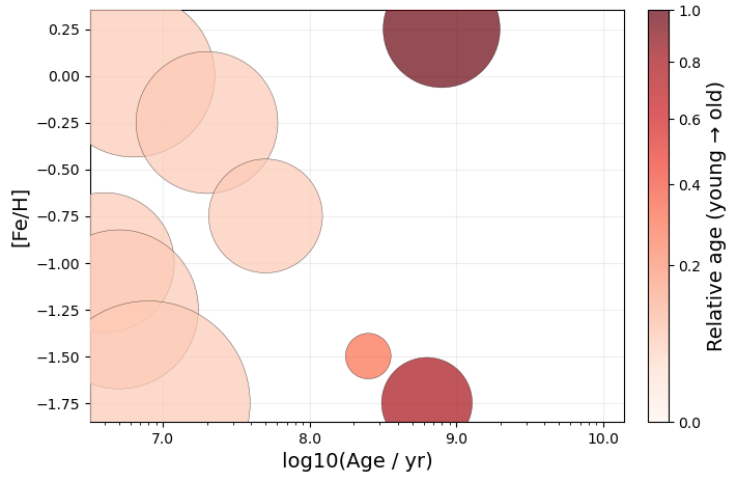}

  \caption{Stellar population properties derived from the \textsc{pPXF} fitting for GAMA~79098 (top row), GAMA~569709 (middle row), and HIPASS J1159–19 S2 (bottom row). \textbf{Left: }Star formation history, shown as the logarithmic stellar mass fraction per Gyr as a function of stellar age (in Gyr). \textbf{Middle:} Stellar metallicity distribution, expressed as stellar mass fraction as a function of [Fe/H]. In this analysis, stars with ages $< 6$\,Gyr are classified as young, while those with ages $> 10$\,Gyr are classified as old. \textbf{Right:} Stellar metallicity as a function of age, where the symbol size encodes the stellar mass fraction and the symbol colour represents stellar age, with older populations shown in redder colours. The distribution curves are smoothed following the binning scale of the stellar-template matrix. Uncertainties on the individual peaks are not shown because the age and metallicity bin widths of the stellar templates exceed the uncertainties in the fitted peak positions.}
  
    \label{fig:stellar}
\end{figure*}

HIPASS J1159–19 S2 is dominated by stellar populations associated with recent or ongoing star formation, with little evidence of a substantial underlying older component (see Fig.~\ref{fig:stellar}). The presence of multiple strongly ionised H\,\textsc{ii} regions further indicates active starburst activity. The system spans a wide range of stellar metallicities, from [Fe/H] $\simeq -2$ to [Fe/H] $\simeq 0$, with a mean stellar metallicity of $\langle\mathrm{[Fe/H]}\rangle = -0.65 \pm 0.02$. Despite being dominated by young stellar populations, HIPASS J1159–19 S2 exhibits an overall metallicity that is comparable to, or marginally lower than, those of GAMA~79098 and GAMA~569709.

The robustness of the \textsc{pPXF} SPS fitting was tested by introducing small variations to the galaxy spaxel mask, spectral range, and regularisation used for the fitting. These changes do not significantly affect the derived stellar population distribution, with the recovered stellar age distributions remaining largely consistent across all tests. The metallicity estimates for the GAMA dwarfs are comparatively less reliable than those for HIPASS~J1159--19~S2 due to their lower continuum SNR, which results in a less reliable tracking of key absorption features and larger uncertainties in the metallicity measurements.

\subsection{Gas-Phase Properties}
\subsubsection{Emission Characteristics}
\paragraph{GAMA~79098} exhibits prominent nebular emission features. The strongest detected lines include H$\alpha$, H$\beta$, H$\gamma$, [O\,\textsc{ii}] $\lambda\lambda3727,3729$, [O\,\textsc{iii}] $\lambda\lambda4959,5007$, [N\,\textsc{ii}] $\lambda\lambda6548,6583$, and [S\,\textsc{ii}] $\lambda\lambda6717,6731$. The internal dust attenuation, parameterised by the colour excess $E(B-V)$, is measured to be $0.15 \pm 0.013$. The gas-phase metallicity, expressed as the oxygen abundance $12 + \log(\mathrm{O/H})$ and derived from the spectrum integrated over star-forming spaxels, is $8.00 \pm 0.08$. The SFR inferred from the H$\alpha$ luminosity using the spectrum integrated over all galaxy spaxels, is $0.0169 \pm 0.002~M_\odot~\mathrm{yr}^{-1}$, with more than 80\% of the total H$\alpha$ emission originating from the central ionised region (see Fig.~\ref{fig:observation}).

\paragraph{GAMA~569709} exhibits only weak gas-phase emission features, indicative of a severely depleted ionised interstellar medium. A faint H$\alpha$ emission line is detected, accompanied by weak [O\,\textsc{iii}] emission, while the H$\beta$ and H$\gamma$ lines are dominated by strong stellar absorption. The absence of prominent metal-line emission precludes a detailed characterisation of the ionised gas component. As the galaxy spectrum is dominated by stellar absorption features, reliable estimates of the internal colour excess or gas-phase metallicity cannot be obtained using standard nebular diagnostics. The inferred SFR for GAMA~569709 is $(1.9 \pm 0.2)\times 10^{-4}~M_\odot~\mathrm{yr}^{-1}$, with the detected H$\alpha$ emission spatially confined to a compact central region. Although the SPS fitting indicates the presence of a young stellar component, the paucity of ionised gas suggests that star formation has largely ceased, consistent with a post-starburst dwarf galaxy \citep{Wenhao_2023}. This behaviour, together with the early-type dwarf morphology of GAMA~569709, suggests that the system may represent a transition-type dwarf between dwarf irregular and dwarf elliptical.

\paragraph{HIPASS J1159–19~S2} is a dwarf irregular galaxy characterised by numerous strongly ionised H\,\textsc{ii} regions embedded within an extended stellar body that is otherwise largely devoid of detectable nebular emission. These ionised regions exhibit strong emission in all major diagnostic lines, including H$\alpha$, H$\beta$, H$\gamma$, [O\,\textsc{ii}] $\lambda\lambda3727,3729$, [O\,\textsc{iii}] $\lambda\lambda4959,5007$, [N\,\textsc{ii}] $\lambda\lambda6548,6583$, and [S\,\textsc{ii}] $\lambda\lambda6717,6731$, indicating ongoing starburst activity within the nebulae. For the spectrum integrated over the star-forming spaxels, the measured colour excess is $E(B-V) = 0.18 \pm 0.07$, and the gas-phase metallicity is $12 + \log(\mathrm{O/H}) = 8.14 \pm 0.15$. The total SFR of the galaxy is estimated to be $0.0476 \pm 0.006~M_\odot~\mathrm{yr}^{-1}$, with approximately 85\% of the H$\alpha$ flux originating from the ionised nebular regions.

\subsubsection{BPT Analysis}
\begin{figure*}

  \includegraphics[width=0.45\linewidth]{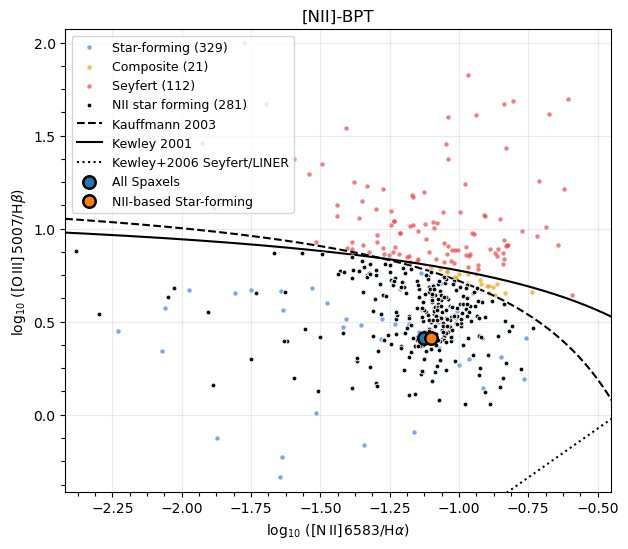}
  \includegraphics[width=0.45\linewidth]{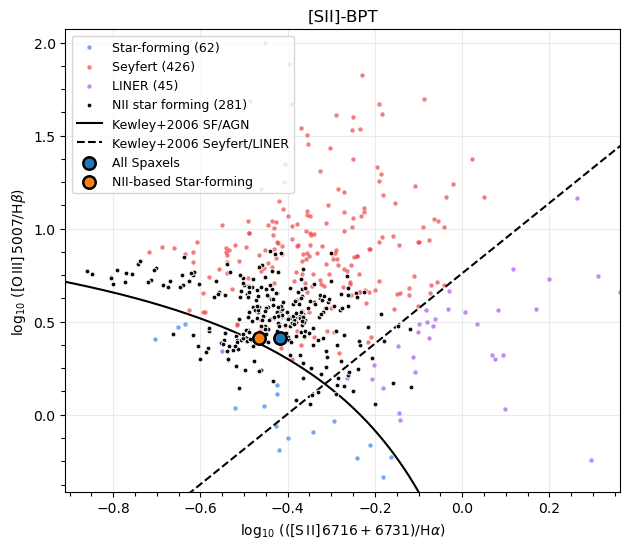}
  \\
  
  \includegraphics[width=0.45\linewidth]{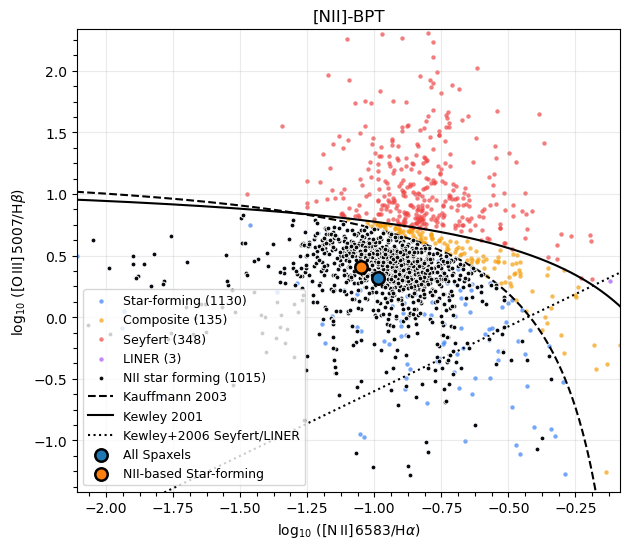}
  \includegraphics[width=0.45\linewidth]{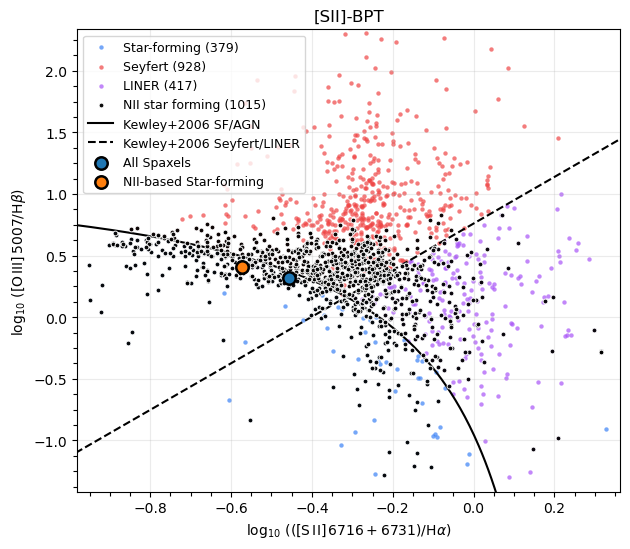}

  \caption{BPT diagrams showing the emission-line classifications (i.e.\ star-forming, composite, Seyfert, and LINER) for individual spaxels in the two dwarf galaxies GAMA~79098 (top row) and HIPASS~J1159$-$19~S2 (bottom row). The large orange and blue filled circles indicate the BPT locations derived from spectra integrated over all galaxy spaxels and over spaxels classified as star-forming based on the [N\,\textsc{ii}]--BPT diagnostic and H$\alpha$ emission, respectively. Small blue, yellow, red, and purple points correspond to spaxels classified as star-forming, composite, Seyfert, and LINER according to the respective BPT demarcation curves. Small black markers denote spaxels identified as star-forming from the [N\,\textsc{ii}] diagnostic. The BPT classification curves are adopted from \citet{Kewley_2001}, \citet{Kauffmann_2003}, and \citet{Kewley_2006}.}
  \label{fig:BPT}
\end{figure*}

\begin{figure*}

  \includegraphics[width=0.4\linewidth, height = 0.4\linewidth]{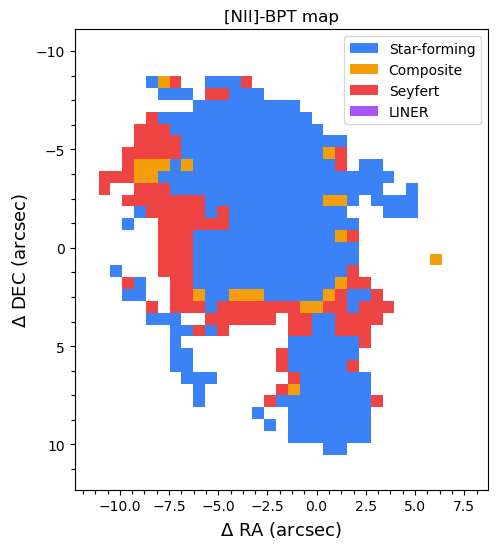}
  \includegraphics[width=0.4\linewidth, height = 0.4\linewidth]{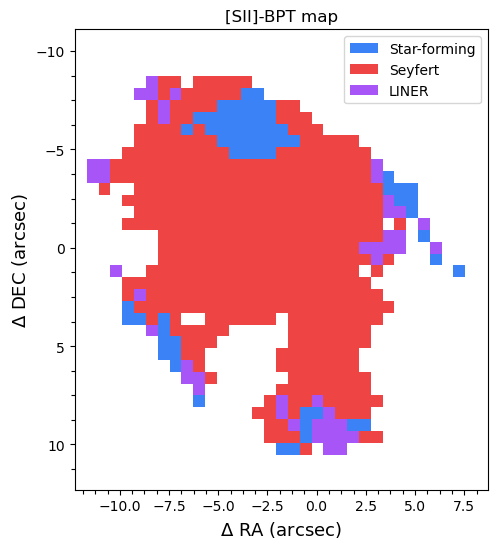}
  \\
  
  \includegraphics[width=0.4\linewidth, height = 0.4\linewidth]{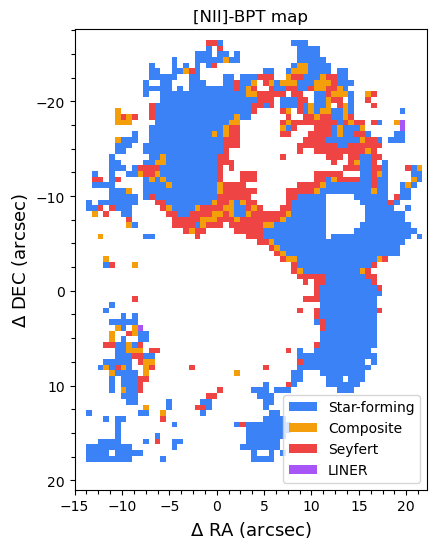}
  \includegraphics[width=0.4\linewidth, height = 0.4\linewidth]{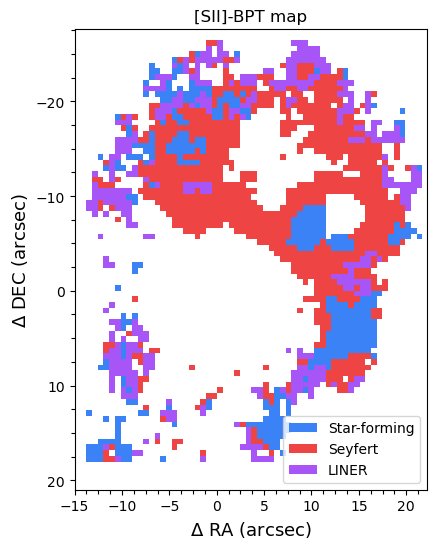}

  \caption{2D pixel map showing the BPT ionisation classification for the spaxals with sufficient H$\alpha$, H$\beta$, [O\textsc{iii}]5007, [N\textsc{ii}]6583, [S\textsc{ii}]6716,6731 line emission, for the galaxies GAMA~79098 (top row) and HIPASS~J1159-19~S2 (bottom row). Left column shows the classification using [N\textsc{ii}]-line based diagnostic. The right column shows the classification using [S\textsc{ii}]-line based diagnostic. Spaxals classified as ionised by star-forming, composite, Seyfert and LINER are coloured blue, yellow, red and purple, following the colouring scheme from previous BPT plot.}
  \label{fig:BPT_map}
\end{figure*}

The BPT diagrams for GAMA~79098 and HIPASS~J1159$-$19 S2 are presented in Fig.~\ref{fig:BPT}, with their spatial distribution displayed in Fig.~\ref{fig:BPT_map}. As shown, the [N\,\textsc{ii}]–BPT diagram classifies both galaxies as primarily ionised by star formation, whereas the [S\,\textsc{ii}]–BPT diagram places them closer to the boundary between the star-forming and AGN regimes. In particular, GAMA~79098 would be classified as AGN-ionised if one relies solely on the [S\,\textsc{ii}] diagnostic.

Similar discrepancies between the [N\,\textsc{ii}] and [S\,\textsc{ii}] diagnostics in dwarf galaxies have been reported in several studies searching for dwarf AGN candidates \citep{Mezcua_2024, Heckler_2024}. In the low-stellar-mass regime, the interstellar medium is typically metal-poor, which can significantly affect emission-line ratios in narrow-line regions. As metallicity decreases, the nitrogen abundance declines relative to hydrogen, leading to systematically lower [N\,\textsc{ii}]/H$\alpha$ ratios. Consequently, data points shift towards the left-hand side of the [N\,\textsc{ii}]–BPT diagram and may occupy the star-forming locus even in the presence of a low-luminosity AGN \citep{Kewley_2006, Reines_2013, Cann_2019, Heckler_2024}.

Low-luminosity Seyfert nuclei have indeed been identified in dwarf galaxies \citep{Filippenko_1989, Peterson_2005, Baldassare_2015}. An example is the irregular dwarf galaxy J1329+3234, which has a stellar mass of $2.0\times10^{8}M_\odot$ and a metallicity of approximately $0.4,Z_\odot$, comparable to that of the SMC and our target dwarfs \citep{Secrest_2015}. However, the current BPT results alone are insufficient to argue that GAMA~79098 is a potential AGN candidate. The [S\,\textsc{ii}]--BPT classification based on the galaxy-integrated spectrum lies very close to the diagnostic curve boundary and shows strong disagreement with the [N\,\textsc{ii}] diagnostic. Moreover, the observed ionisation levels could be influenced by alternative mechanisms such as shocks or hard stellar radiation fields \citep{Oparin_2018}.

To further assess the presence of AGN activity, we examined available X-ray and infrared (IR) survey data for both GAMA~79098 and HIPASS~J1159$-$19~S2. From the XMM--Newton Slew Survey\footnote{\url{https://sci.esa.int/web/xmm-newton/-/43133-slew-survey-and-catalogue}}, there are no reported significant X-ray detections associated with either system. We additionally considered mid-IR colours from the Wide-field Infrared Survey Explorer (WISE)\footnote{\url{https://irsa.ipac.caltech.edu/data/WISE/docs/release/All-Sky/}}, specifically the $W1 - W2$ colour, where $W1$ and $W2$ correspond to the 3.4 and 4.6~$\mu$m bands, respectively. Following the AGN selection criteria of \citet{Stern_2012} and \citet{Assef_2013}, galaxies with $W2 < 15.05$ and $W1 - W2 > 0.8$ are classified as AGN candidates. Both GAMA~79098 ($W2 = 14.64 \pm 0.06$) and HIPASS~J1159$-$19~S2 ($W2 = 14.09 \pm 0.05$) have $W2$ magnitude less than 15.05; however, their measured colours are significantly below the AGN threshold, with $W1 - W2 = 0.12 \pm 0.07$ and $0.27 \pm 0.06$, respectively. Therefore, neither the available X-ray nor mid-IR diagnostics provide supporting evidence for AGN activity in GAMA~79098 or HIPASS~J1159$-$19~S2.

\subsection{Velocity Fields}

\begin{figure*}

  \includegraphics[width=0.31\linewidth, height = 0.4\linewidth]{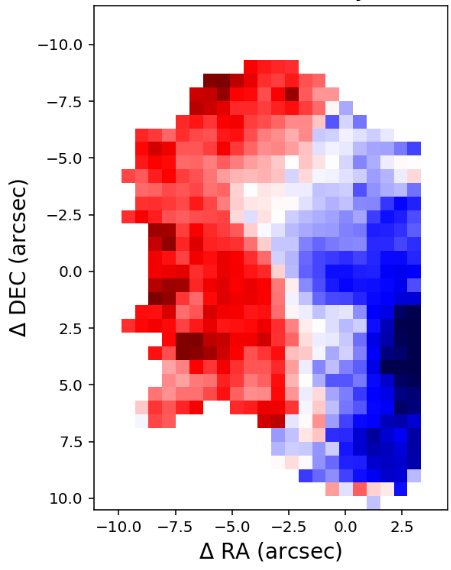}
  \includegraphics[width=0.29\linewidth, height=0.4\linewidth]{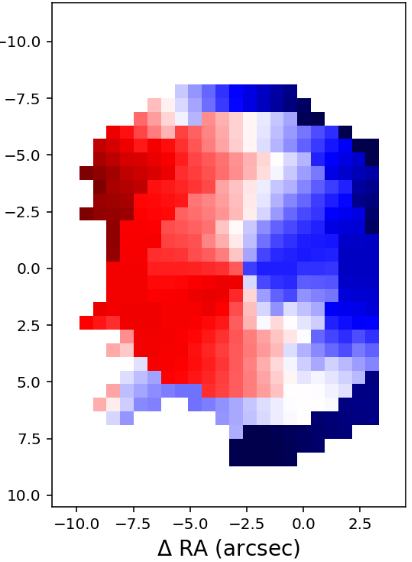}
  \includegraphics[width=0.36\linewidth, height=0.4\linewidth]{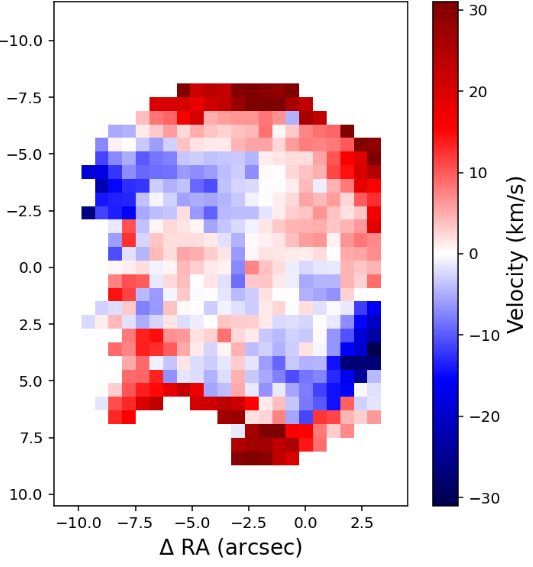}

  \caption{\textbf{Left:} Line-of-sight velocity map of GAMA~79098 derived from the H$\alpha$ emission-line peak position in each spaxel. Red colours indicate velocities redshifted relative to the median H$\alpha$-based systemic velocity of the galaxy, while blue colours indicate blueshifted velocities. Middle: Model velocity field obtained from two-dimensional harmonic ring fitting. The outermost ring is visibly affected by overfitting, primarily due to the limited number of spaxels sampling certain azimuthal angles. \textbf{Right:} Residual velocity map obtained by subtracting the model velocity field from the observed line-of-sight velocity distribution (LOSVD).}
  \label{fig:GAMA79098_vfield}
\end{figure*}

\paragraph{GAMA~79098} LOSVD analysis is illustrated in Figure~\ref{fig:GAMA79098_vfield}. The measured velocity shear and the observed velocity dispersion are $V_S = 20.4 \pm 1.8$\,\kms and $\sigma_{V} = 14.5 \pm 0.8$\,\kms, respectively. Harmonic fitting of the velocity field reveals a clear bulk rotational component. The global kinematic position angle (PA) of this rotation is $13 \pm 8$ degrees measured counter-clockwise relative to the horizontal axis. The corresponding rotation amplitude is $V_{\mathrm{rot}} = 20.2 \pm 1.9$\,\kms.

\begin{figure}

  \includegraphics[width=0.9\columnwidth]{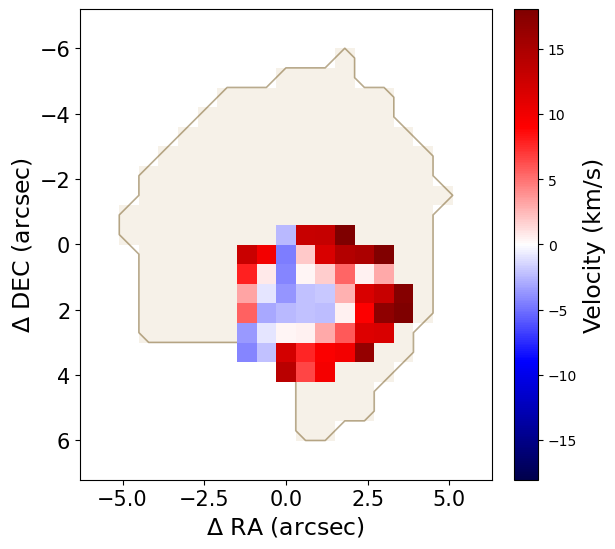}

  \caption{Line-of-sight velocity map of GAMA~569709 derived from the H$\alpha$ emission-line peak position in each spaxel. Red colours indicate velocities redshifted relative to the median H$\alpha$-based systemic velocity of the galaxy, while blue colours indicate blueshifted velocities. The grey background indicates the full extent of the galaxy. The map contains only a limited number of spaxels because H$\alpha$ emission is detected only in the central region. The bottom left corner of the galaxy is masked out due to a foreground star.}
  \label{fig:GAMA569709_vfield}
\end{figure}

\paragraph{GAMA~569709} 
LOSVD map shows no significant evidence for bulk rotation (see Fig.~\ref{fig:GAMA569709_vfield}), at least within the relatively compact region characterised by strong H$\alpha$ emission (SNR $> 5$). Within this region, the measured velocity shear and the raw velocity dispersion are $V_S = 10 \pm 5$\,\kms and $\sigma_{V} = 9.4 \pm 1.3$\,\kms, respectively.

\begin{figure*}

  \includegraphics[width=0.31\linewidth, height = 0.4\linewidth]{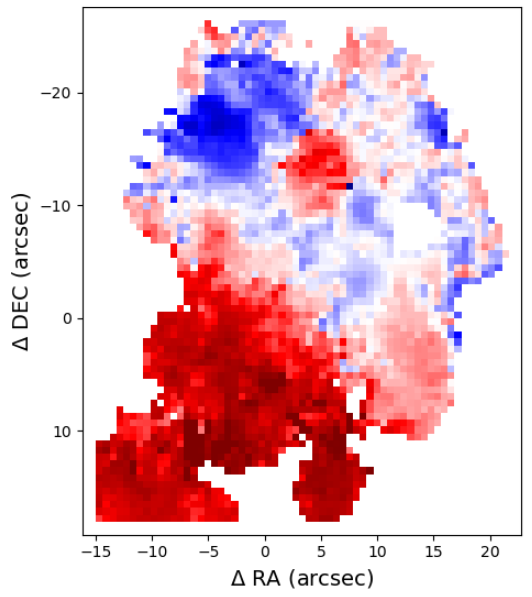}
  \includegraphics[width=0.29\linewidth, height=0.4\linewidth]{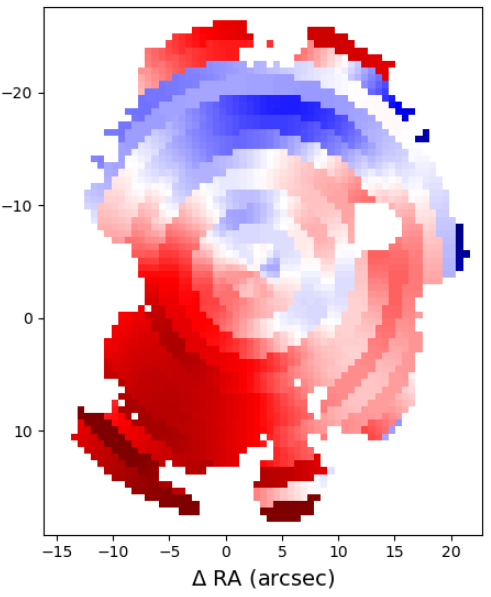}
  \includegraphics[width=0.36\linewidth, height=0.4\linewidth]{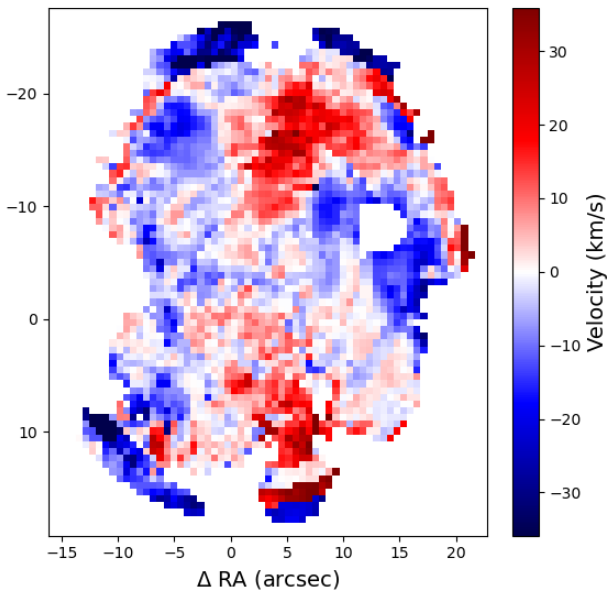}
    
  \caption{\textbf{Left: }Line-of-sight velocity map of HIPASS J1159–19 S2 derived from the peak position of the H$\alpha$ emission line in each spaxel. Red colours denote velocities redshifted relative to the median H$\alpha$-based systemic velocity of the galaxy, while blue colours denote blueshifted velocities. Middle: Model velocity field obtained from two-dimensional harmonic ring fitting. The model reproduces the large-scale velocity discontinuity but does not capture smaller-scale kinematic irregularities. Right: Residual velocity map obtained by subtracting the model velocity field from the observed line-of-sight velocity distribution. Small-scale, kinematically irregular gas clouds that do not participate in the large-scale rotational pattern are visible in the residual map.}
    \label{fig:HIPASS_vfield}
\end{figure*}

\paragraph{HIPASS J1159–19~S2} 
LOSVD map is presented in 
Figure~\ref{fig:HIPASS_vfield}, computed over regions with sufficient H$\alpha$ SNR. The measured velocity shear and the raw velocity dispersion are $V_S = 26.9 \pm 0.8$\kms and $\sigma_{v} = 17.2 \pm 0.6$\,\kms, respectively. The kinematic PA derived from the harmonic fitting is $79 \pm 6.8$ degrees. The maximum rotation amplitude of the harmonic model fit is $V_{\mathrm{rot}} = 22.6 \pm 1.2$\,\kms. The H\,\textsc{i} LOSVD map reported by \citet{Phookun_1992}, covering a wide region around HIPASS J1159$-$19 S1 and S2 (See their Fig. 2), reveals rough velocity discontinuities across the region. Although the H\,\textsc{i} velocity field exhibits only a slight gradient at the location of HIPASS J1159-19 S2, the presence of large-scale discontinuities in the H\,\textsc{i} velocity field suggests that the prominent velocity discontinuity seen in our LOSVD is a genuine physical feature.

The median rotation velocity, $18.4 \pm 1.2$\,\kms, might provide a more representative measure of the global rotation of HIPASS J1159–19 S2, as the maximum rotation amplitude may be overestimated due to the kinematic influence of individual H\,\textsc{ii} regions. Both the LOSVD and residual velocity maps indicate that prominent H\,\textsc{ii} clumps possess kinematic properties distinct from the surrounding stellar body. The observed kinematic heterogeneity, characterised by irregular motions of the inner ionised regions and a pronounced velocity discontinuity, suggests the possibility of a significant tidal interaction with the nearby massive galaxy NGC 4027 or a merger with a nearby dwarf.

\begin{table}

\begin{tabular}{@{}rcccc@{}}
\hline
\multicolumn{1}{c}{}                 & Units                                      & \begin{tabular}[c]{@{}c@{}}GAMA \\ 79098\end{tabular} & \begin{tabular}[c]{@{}c@{}}GAMA \\ 569709\end{tabular} & \begin{tabular}[c]{@{}c@{}}HIPASS \\ J1159-19 S2\end{tabular} \\ \hline
\multicolumn{1}{|r|}{M$_r$ (1)}      & \multicolumn{1}{c|}{mag}                   & $-16.5\pm$0.4                                         & $-14.4\pm$0.5                                          & $-17.9\pm$0.6                                                 \\
\multicolumn{1}{|r|}{M$_*$ (2)}      & \multicolumn{1}{c|}{dex(M/M$_\odot$)}      & 8.5$\pm$0.2                                           & 8.0$\pm$0.3                                            & 7.8$\pm$0.3                                                   \\
\multicolumn{1}{|r|}{[Fe/H]$_*$ (3)} & \multicolumn{1}{c|}{-}                     & $-0.7\pm$0.3                                          & $-0.7\pm$0.4                                           & $-0.65\pm$0.02                                                \\
\multicolumn{1}{|r|}{Z$_{gas}$ (4)}  & \multicolumn{1}{c|}{12+log(O/H)}           & 8.00$\pm$0.08                                         & -                                                      & 8.14$\pm$0.15                                                 \\
\multicolumn{1}{|r|}{SFR (5)}        & \multicolumn{1}{c|}{10$^{-2}$M$_\odot$/yr} & 1.69$\pm$0.02                                         & 2$\pm$0.2($\times$10$^{-2}$)                           & 4.76$\pm$0.06                                                 \\
\multicolumn{1}{|r|}{\textit{E(B-V)} (6)}   & \multicolumn{1}{c|}{-}                     & 0.15$\pm$0.013                                        & -                                                      & 0.18$\pm$0.07                                                 \\
\multicolumn{1}{|r|}{V$_{shear}$(7)} & \multicolumn{1}{c|}{\kms}                  & 20.4$\pm$1.8                                          & 10$\pm$5                                               & 26.9$\pm$0.8                                                  \\
\multicolumn{1}{|r|}{$\sigma_V$(8)}  & \multicolumn{1}{c|}{\kms}                  & 14.5$\pm$0.8                                          & 9.4$\pm$1.3                                            & 17.2$\pm$0.6                                                  \\
\multicolumn{1}{|r|}{V$_{rot}$ (9)}  & \multicolumn{1}{c|}{\kms}                  & 20.2$\pm$1.9                                          & -                                                      & 22.6$\pm$1.2                                                  \\
\multicolumn{1}{|r|}{$\epsilon$ (10)}     & \multicolumn{1}{c|}{-}                     & 0.35$\pm$0.02                                         & 0.08$\pm$0.05                                          & 0.29$\pm$0.01                                                 \\
\multicolumn{1}{|r|}{PA (11)}             & \multicolumn{1}{c|}{degrees}               & 134$\pm$2                                             & 105$\pm$27                                             & 70$\pm$1.2                                                    \\ \hline
\end{tabular}

\caption{Summary of the observed properties of the three target galaxies. (1) SDSS $r$-band stellar-only absolute magnitude derived from the \textsc{pPXF} model spectrum; (2) total stellar mass inferred from the \textsc{pPXF} fit; (3) mean stellar metallicity from the \textsc{pPXF} results; (4) gas-phase metallicity of star-forming H\,\textsc{ii} regions measured using equations~\ref{eq:S_U} and~\ref{eq:S_L}; (5) star formation rate derived from galaxy-wide H$\alpha$ emission using equation~\ref{eq:SFR}; (6) $B-V$ colour excess estimated from the H$\gamma$/H$\beta$ flux ratio; (7) velocity shear measured from the H$\alpha$-based LOSVD (see equation~\ref{eq:V_S1}); (8) LOSVD velocity dispersion (see equation~\ref{eq:V2}); (9) galaxy rotation amplitude derived from harmonic ring modelling (see equation~\ref{eq:gal_v}); (10) galaxy ellipticity derived from the galaxy spaxel mask; (11) galaxy position angle derived from the same mask.}
\label{tab:results}
\end{table}


\section{Discussion}
In this section, we interpret the observed stellar, gas-phase, and kinematic properties of the target dwarf galaxies in the context of Local Group analogues and established scaling relations. As the Delegate Survey aims to assess the typicality of the LG, we study the degree to which the analysed LGA dwarfs resemble their LG counterparts, with particular emphasis on similarities and differences in star formation histories, chemical enrichment, and dynamical structure. Where appropriate, we discuss the potential role of the environment and evolutionary stage in shaping the observed properties, while considering the limitations imposed by the sample size and observational methodology.

The absolute magnitudes and corresponding stellar mass estimates reported for LG dwarf galaxies are generally derived from observations that have not been corrected for internal dust extinction. To ensure a consistent comparison, we therefore adopt absolute magnitudes and stellar masses for our LGA dwarfs derived from spectra that are likewise uncorrected for internal extinction, applying corrections only for foreground MW extinction. In contrast, comparisons of stellar and gas-phase metallicities are performed using values derived from fully extinction-corrected spectra, as previous studies of H\,\textsc{ii} region and planetary nebula metallicities in LG dwarfs have consistently applied internal extinction corrections.

Moreover, the majority of LG dwarf galaxy properties available in the literature are based on photometric measurements or long-slit and single-fibre spectroscopy, which are inherently subject to aperture biases. As a result, differences between our LGA dwarf sample and LG dwarfs may partly reflect systematic effects arising from the distinct observational techniques employed. In addition, our kinematic measurements are based on H$\alpha$ emission tracing ionised gas within H\,\textsc{ii} regions, which may not directly correspond to kinematics derived from stellar or H\,\textsc{i} components, and therefore provide only an approximate representation of the overall dynamical state. To remain robust against these methodological differences, we consider galaxies to be broadly comparable when their properties agree within approximately a factor of two. Future expansion of the Delegate sample and uniform analysis of both LG and LGA dwarfs will enable more precise and direct comparisons, mitigating these systematic uncertainties.

\subsection{Comparison Sample }
Here we describe the properties of a subset of LG dwarf galaxies selected for comparison with our LGA dwarf sample. The comparison galaxies are drawn from the MW and M31 subgroups, reflecting the fact that all of our target galaxies are located in close proximity to central host galaxies. We select LG dwarfs with absolute $V$-band magnitudes in the ranges $-13.75 \lesssim M_V \lesssim -15.15$ and $-15.75 \lesssim M_V \lesssim -18.65$, chosen to bracket the SPS-modelled absolute $V$-band magnitudes of our targets within a factor of two in luminosity. These ranges are defined such that the absolute magnitudes of the LG comparison sample fall within the uncertainty bounds of the $V$-band model magnitudes derived for the LGA dwarfs from the \textsc{pPXF} fitting.

Five LG dwarf galaxies satisfy these selection criteria, and their properties are summarised in Table~\ref{tab:LG_dwarfs}. The majority of the data are taken from the compilation of observed LG dwarf properties presented by \citet{McConnachie_2012}. The properties of target LG dwarfs not listed in \citet{McConnachie_2012} are supplemented where necessary with values from the following studies. The galaxy size estimator is given as $B$-band $D_{25}$ major-axis diameter (25\,mag\,arcsec$^{-2}$ isophote) from \cite{deVacucoulurs_1991}. Gas-phase metallicity estimates for LMC and SMC are derived from emission-line ratios measured in ionised H\textsc{ii} regions \citep{Cipriano_2017}. For the remaining dwarf galaxies, which lack large ionised star-forming regions, we adopt error-weighted mean metallicities derived from planetary nebulae as a proxy for gas-phase abundances \citep{McCall_2003, Gonçalves_2007, Magrini_2009}.

\begin{table*}

\centering
    \begin{tabular}{cccccccccc}
    \hline
        Name & Morp & $M_V$& $M_*$& $\Delta R_{cent}$ & $D_{25}$ & $v_{rot}$ & $\sigma_V$ & $\langle Z_*\rangle$ & $\langle Z_{gas}\rangle$ \\
        Units & - & mag &$10^6M_\odot$ & kpc & kpc & \kms & \kms & $\langle $\feh$ \rangle$  & $12+\rm{log}(O/H)$  \\
        (1) & (2) &(3)&  (4)& (5) & (6)&  (7)&(8)& (9) & (10)\\
        \hline
         LMC & dIrr & $-18.1$ & 2700 & 50 & 9.3 & $49.8\pm15.9$ & $20.2\pm0.5$ &  $-0.5$ & $8.35\pm0.03$ \\
       SMC & dIrr & $-16.8$ & 460 & 61 & 5.7 & $20\pm0.7$ & $27.6\pm0.5$ & $-1.00\pm0.02$ & $8.03\pm0.03$  \\
        NGC 205 & dE/dSph & $-16.5$ & 330 & 42 & 5.23 & $11\pm5$ & $35\pm5$ & $-0.8\pm0.2$ & $8.46\pm 0.02$ \\
        NGC 185 & dE/dSph & $-14.8$ & 68 & 187 & 2.12 & $15\pm5$ & $24\pm1$ & $-1.3\pm0.1$ & $7.99\pm0.03$ \\
        NGC 147 & dE/dSph & $-14.6$ & 62 & 142 & 2.59 & $17\pm2$ & $16\pm1$ & $-1.1\pm0.1$ & $8.0\pm 0.1$ \\
        \hline
    \end{tabular}
    \caption{Basic properties of five LG dwarf galaxies. (1) Galaxy name; (2) morphology; (3) $V$-band absolute magnitude \citep{deVacucoulurs_1991, Choi_2002, McConnachie_2006}; (4) stellar mass estimate \citep{vanDerMarel_2002, Harris_2006, Geha_2010}; (5) distance to the nearest massive host (MW or M31) \citep{Udalski_1999, vanDerMarel_2002, Clementini_2003,McConnachie_2005, Harris_2006, Geha_2006b, Geha_2010}; (6) $B$-band $D_{25}$ major-axis diameter (25\,mag\,arcsec$^{-2}$ isophote) from \citet{deVacucoulurs_1991}; (7) maximum rotational velocity measured from stellar kinematics corrected for inclination \citep{vanDerMarel_2002, Geha_2006b, Geha_2010, Piatti_2021}; (8) velocity dispersion from stellar kinematics \citep{vanDerMarel_2002, Geha_2006b}; (9) mean stellar metallicity \citep{McConnachie_2005, Geha_2006b, Carrera_2008, Geha_2010, Parisi_2010}; (10) gas-phase metallicity estimate. Values are primarily taken from the Local Group galaxy compilation of \citet{McConnachie_2012}, with the original references given above. For the SMC and LMC, gas-phase metallicities are derived from emission-line ratios in ionised H\,\textsc{ii} regions \citep{Cipriano_2017}. For NGC~205, NGC~185, and NGC~147, we report error-weighted mean metallicities derived from planetary nebulae \citep{McCall_2003, Gonçalves_2007}. The LMC stellar mass is updated using the value reported by \citet{Shipp_2021}, and the SMC rotation velocity is adopted from \citet{Zivick_2021}.}
    
    \label{tab:LG_dwarfs}
\end{table*}

\subsubsection{GAMA~79098}
GAMA~79098 shows a close correspondence with the SMC in terms of luminosity, stellar mass, kinematic properties, and chemical abundances, and can therefore be regarded as a lower-mass analogue of the SMC. The $V$-band absolute magnitude of GAMA~79098 derived from the SPS fitting is $M_V = -16 \pm 0.5$, and its stellar mass is estimated to be $(2.5 \pm 0.9)\times 10^8\,M_\odot$. These values are consistent within a factor of two with those of the SMC ($M_V = -16.8$, $M_\star = 4.6 \times 10^8\,M_\odot$). GAMA~79098 also exhibits a kinematic profile characterised by $V_{rot} = 20.2 \pm 1.9$\,\kms and $\sigma_V = 14.5 \pm 0.8$\,\kms, where the magnitude of rotation is comparable to that of the SMC ($V_{rot} = 20 \pm 0.7$\,\kms, $\sigma_V = 27.6 \pm 0.5$\,\kms). However, it is important to note that the SMC is subject to significant tidal influence from the MW \citep{Craig_2022}, which may have substantially altered its kinematic properties, thus limiting the reliability of direct kinematic comparisons.

The mean stellar metallicity of GAMA~79098, $\langle \mathrm{[Fe/H]} \rangle = -0.7 \pm 0.3$, and the gas-phase metallicity of its ionised regions, $12 + \log(\mathrm{O/H}) = 8.0 \pm 0.08$, are consistent within the quoted uncertainties with the corresponding values for the SMC ($\langle \mathrm{[Fe/H]} \rangle \sim -1$, $12 + \log(\mathrm{O/H}) \sim 8$). The luminosity-weighted stellar age distribution of GAMA~79098 indicates that approximately 60\% of its stellar mass formed during relatively recent or ongoing starburst episodes, while the remaining $\sim 40$\% formed during the early phases of galaxy assembly at ages $\gtrsim 12$\,Gyr. This ongoing SFH with current strong starburst activity closely resembles that inferred for the SMC, which similarly hosts intense recent star formation superimposed on a substantial old stellar population (approximately 50\% older than $\sim 8$\,Gyr; \citealt{Rubele_2018}).

In addition, GAMA~79098 hosts irregular, highly ionised star-forming complexes with characteristic sizes of several hundred parsecs, embedded within a more diffuse stellar component with low star formation activity. This morphology is analogous to that observed in the SMC, including its prominent NGC~346/N66 star-forming complexes \citep{Gouliermis_2010}, although the star-forming activity of SMC could be significantly affected by interaction with the MW or LMC. Among dwarf irregular galaxies in the MW and M31 subgroups, the SMC is the only system with stellar mass and physical extent comparable to those of GAMA~79098.

Although NGC~205 is classified as an early-type dwarf galaxy, it nonetheless exhibits several similarities to GAMA~79098. NGC~205 has a comparable luminosity and stellar mass  ($M_V = -16.5$, $M_\star = 3.3 \times 10^8\,M_\odot$), as well as a similar mean stellar metallicity ($\langle \mathrm{[Fe/H]} \rangle \sim -0.8$). In contrast, its kinematic structure is more dispersion dominated, with lower rotational velocity and higher velocity dispersion ($V_{rot} = 11 \pm 5$\,\kms; $\sigma_V = 35 \pm 5$\,\kms), and it exhibits a higher gas-phase planetary nebula metallicity ($12 + \log(\mathrm{O/H}) \sim 8.5$). Furthermore, NGC~205 exhibits a distinct star formation history, with the majority of its stellar mass formed at early times and only limited subsequent star formation. In particular, $99.97\%$ of its stellar mass has ages greater than 9.5~Gyr \citep{Leahy_2025}. The present-day star formation rate is $1.3 \times 10^{-3}\,\mathrm{M_\odot\,yr^{-1}}$, which is an order of magnitude lower than that measured for GAMA~79098, distinguishing it from the more extended and ongoing star formation activity observed in GAMA~79098.

\subsubsection{GAMA~569709}
GAMA~569709 shows a close resemblance to NGC~147, with most key properties—including luminosity, stellar mass, and metallicity—being consistent within the quoted uncertainties, while also sharing some similarities with NGC~185. The $V$-band absolute magnitude of GAMA~569709 is measured to be $M_V = -14.3 \pm 0.5$, and its stellar mass is estimated to be $(1.0 \pm 0.5) \times 10^8\,M_\odot$. Both NGC~185 ($M_V = -14.8$, $M_\star = 6.8 \times 10^7\,M_\odot$) and NGC~147 ($M_V = -14.6$, $M_\star = 6.2 \times 10^7\,M_\odot$) exhibit total luminosities and stellar masses that are comparable, within a factor of two, to those of GAMA~569709.

GAMA~569709 exhibits a velocity dispersion of $9.4 \pm 1.3$\,\kms, with no satisfactory evidence for ordered rotation detected in the region with sufficient H$\alpha$ emission. In contrast, both NGC~185 and NGC~147 show higher velocity dispersions of $24 \pm 1$ and $16 \pm 1$\,\kms, respectively, accompanied by measurable internal rotation \citep{Geha_2010}. However, given the limited spatial extent and SNR of the LOSVD measurements for GAMA~569709, a definitive comparison of the kinematic structure cannot yet be made. The mean stellar metallicity of GAMA~569709, $\langle \mathrm{[Fe/H]} \rangle = -0.7 \pm 0.4$, is consistent within uncertainties with that of NGC~147 ([Fe/H]$\sim -1.1$), while NGC~185 exhibits a slightly lower mean stellar metallicity ([Fe/H]$\sim -1.3$).

The derived star formation history of GAMA~569709 indicates that approximately 60\% of its stellar mass formed during an early starburst episode at ages $\gtrsim 12$\,Gyr, with the remaining $\sim 40$\% produced during a more recent starburst that has largely ceased at the present epoch. Studies based on optical photometry and colour--magnitude diagram (CMD) analyses using \emph{Hubble Space Telescope} observations consistently indicate that both NGC~185 and NGC~147 experienced two major episodes of star formation, with their initial starbursts occurring approximately 12.5~Gyr ago \citep{Weisz_2014, Geha_2015}. In NGC~185, the secondary episode took place relatively early, at $\sim$8--10~Gyr ago, whereas star formation in NGC~147 proceeded more gradually, with a subsequent episode occurring at intermediate ages of $\sim$5--7~Gyr ago \citep{Geha_2015}. Photometric CMD-based studies therefore suggest that the SFH of NGC~147 bears a closer resemblance to that of GAMA~569709, characterised by a larger temporal separation between the initial and secondary star-forming episodes, although both systems show evidence for early quenching.

In contrast, SFH estimates for NGC~185 derived from stellar population synthesis (SPS) modelling suggest that a substantial fraction of its stellar luminosity (approximately $30\%$) is from stars with younger age ($<500$~Myr) \citep{Martins_2012}, yielding a closer similarity to the SFH inferred for GAMA~569709. This apparent discrepancy may arise because younger stellar populations are intrinsically brighter and thus exert a disproportionate influence on the integrated galaxy spectra. Consequently, it remains plausible that GAMA~569709 possesses an SFH more closely aligned with that of NGC~185, and potentially also NGC~147, which itself exhibits a high degree of similarity to NGC~185.

Furthermore, GAMA~569709 lies at a projected separation of 127.8~kpc from its nearest massive spiral galaxy, comparable to projected separations of 142~kpc and 95~kpc between M31 and NGC~147 and NGC~185, although the true three-dimensional separation of GAMA~569709 could be significantly greater. Given that GAMA~569709 appears to be transitioning towards a quenched, early-type system, it may plausibly evolve into a galaxy analogous to NGC~147 or NGC~185 on Gyr timescales.

\subsubsection{HIPASS J1159-19 S2}
HIPASS J1159–19 S2 exhibits luminosity, kinematic properties, and chemical abundances that are intermediate between those of the SMC and the LMC. The SPS-modelled $V$-band absolute magnitude of HIPASS J1159–19 S2 is $M_V = -17.5 \pm 0.8$, and its stellar mass is estimated to be $( 1\pm 0.4) \times 10^8\,M_\odot$. Within this uncertainty range, the SMC ($M_V = -16.8$, $M_\star = 4.6 \times 10^8\,M_\odot$) falls within a factor of two, whereas the LMC exhibits a comparable luminosity but a substantially higher stellar mass ($M_V = -18.1$, $M_\star = 2.7 \times 10^9\,M_\odot$; \citealt{Shipp_2021}).

The kinematic properties of HIPASS J1159–19 S2, characterised by $V_{rot} \sim 22.6$\,\kms and $\sigma_V \sim 17.2$\,\kms, closely resemble those of the SMC ($V_{rot} \sim 20$\,\kms; $\sigma_V \sim 28$\,\kms). In contrast, the LMC displays a significantly higher rotation velocity of $50 \pm 16$\,\kms, while maintaining a comparable velocity dispersion of $20.2 \pm 0.5$\,\kms. The mean stellar metallicity of HIPASS J1159–19 S2, $\langle \mathrm{[Fe/H]} \rangle = -0.65 \pm 0.02$, lies between the characteristic values of the LMC ($\sim -0.5$\,dex) and the SMC ($\sim -1$\,dex), remaining comparable within a factor of two (approximately 0.3\,dex). Similarly, the gas-phase metallicity of HIPASS J1159–19 S2, $12 + \log(\mathrm{O/H}) = 8.14 \pm 0.15$, is intermediate between the values measured for the LMC ($8.35 \pm 0.03$) and the SMC ($8.03 \pm 0.03$).

The luminosity-weighted stellar age distribution of HIPASS J1159–19 S2 indicates that the vast majority of its stellar mass (approximately 90\%) is composed of young stars with ages younger than 2\,Gyr. Only a minor fraction of older, relatively metal-rich stars with ages of order $\sim 6$\,Gyr is present. As a member of the Choir group, HIPASS J1159–19 is associated with a high H\textsc{i} content and is thought to be at an early stage of group assembly \citep{Sweet_2013}. In contrast, both Magellanic Clouds host substantial old stellar populations formed during the early starburst phases of galaxy assembly \citep{Harris_2009, Rubele_2018}.

HIPASS J1159–19 S2 also contains irregular, strongly ionised star-forming complexes with characteristic sizes of several hundred parsecs embedded within regions of relatively low global star formation activity, resembling the prominent H\,\textsc{ii} regions observed in the SMC and LMC, such as NGC~346/N66 and NGC~2070 \citep{Vacca_1995, Gouliermis_2010}. Furthermore, HIPASS J1159–19 S2 lies at a projected separation of 28.06\,kpc from its host galaxy NGC~4027, comparable to the separation between the SMC and the LMC. Moreover, its close proximity to the host galaxy suggests that its star formation may have been influenced by tidal interactions. This interpretation is supported by the location of HIPASS J1159--19 S2 within the disturbed H{\sc i} velocity field surrounding HIPASS J1159--19 \citep{Phookun_1992}.
 Taken together, these properties suggest that HIPASS J1159–19 S2 may represent a newly forming massive dwarf irregular galaxy, potentially analogous to an early evolutionary stage of the SMC or LMC.

\subsubsection{Mass-Metallicity Relation}
\begin{figure}

  \centering
  \includegraphics[width=\columnwidth]{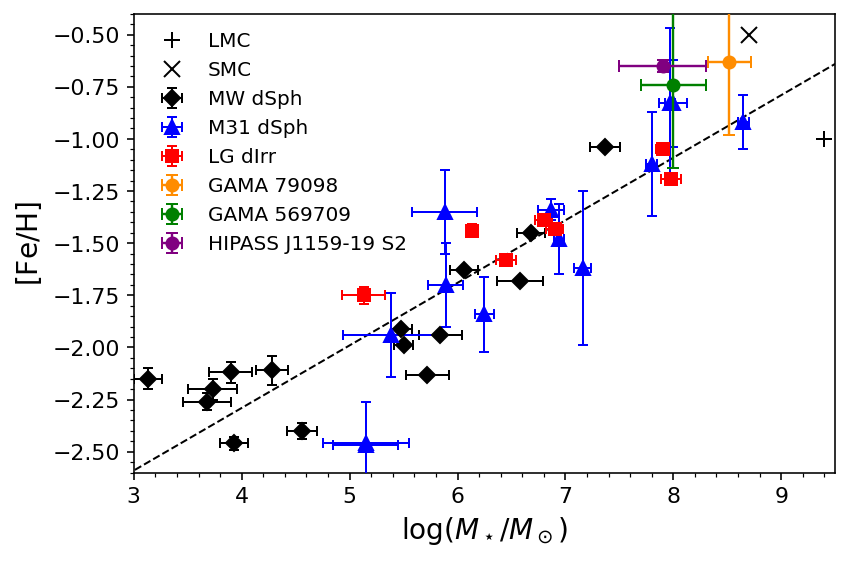}
 
  \caption{Stellar mass and stellar metallicity of the three target dwarf galaxies overlaid on the stellar mass–metallicity relation (MZR) for Local Group dwarfs from \citet{Kirby_2013}. Measurement uncertainties are indicated by the horizontal and vertical error bars. The dashed line represents the least-squares fit to the Local Group dwarf sample, excluding the LMC and SMC, as presented in \citet{Kirby_2013}. Data points for the LMC and SMC are added based on the values listed in Table~\ref{tab:LG_dwarfs}.
 }
   \label{fig:LGA_MZR}
\end{figure}

The results for the three target galaxies are overlaid on the stellar mass–metallicity relation (MZR) of LG dwarf galaxies from \citet{Kirby_2013} in Fig.~\ref{fig:LGA_MZR}. The two GAMA group dwarfs follow a similar linear trend and are consistent with the LG MZR within the reported uncertainties, although they may tentatively indicate slightly higher stellar metallicities at fixed stellar mass. However, given the small sample size and the limited stellar mass range probed, no robust conclusion can be drawn regarding systematic differences between the GAMA group and LG dwarf MZRs. Moreover, \citet{Kirby_2013} derive stellar metallicities from the Fe\,\textsc{i} line spectral synthesis of individual stars in LG dwarf galaxies, rather than from integrated spectra using \textsc{pPXF}, which further limits the direct comparability of the samples. Future observations and analysis of the remaining dwarf galaxies in the GAMA J1440-00 group will enable a more robust comparison by extending the sample across a broader stellar-mass range and including a larger number of low-mass systems.

In contrast, HIPASS J1159–19 S2 exhibits a significantly higher metallicity at its stellar mass compared to LG dwarf galaxies of similar mass, with the offset exceeding the associated uncertainty bounds. Given the relatively high SNR of HIPASS~J1159--19~S2, and the correspondingly more reliable stellar mass and metallicity estimates, the observed offset from the MZR appears genuine. This behaviour is qualitatively consistent with expectations, as the HIPASS group is likely dynamically younger and exhibits distinct physical and evolutionary characteristics from the dwarfs in the LG \citep{Sweet_2013}, potentially leading to differences in the chemical enrichment histories.

\section{Conclusions}
This work presents the first scientific results of the Delegate Survey through integral-field spectroscopic observations of three luminous dwarf galaxies -- GAMA~79098, GAMA~569709, and HIPASS~J1159--19~S2 -- obtained using the KOALA integral-field unit coupled with the AAOmega spectrograph on the AAT. The targets were selected from the Local Group Analogue group GAMA~J1440--00 and from the Choir group HIPASS~J1159--19. They were chosen based on their relatively high surface brightness to validate observational strategies and data-analysis methodology.

In this work, a comprehensive reduction and analysis workflow was developed and tested. The raw data were processed using \textsc{2dfdr}, followed by cube reconstruction and calibration with \textsc{PyKOALA}. Stellar population properties were derived using \textsc{pPXF} full-spectrum fitting on the blue arm, while emission-line fluxes were measured from continuum-subtracted spectra for spatially resolved diagnostics. The development and validation of this workflow constitute a foundational technical outcome of this paper and establish the methodological framework for the broader Delegate Survey.\\

Our principal scientific results can be summarised as follows:
\begin{enumerate}
    \item 
    GAMA~79098 and GAMA~569709 both host composite stellar populations with an early major star-forming episode ($10$--$14$\,Gyr ago). GAMA~79098 continues to exhibit strong ongoing star formation, with $\gtrsim60\%$ of its stellar mass formed within the past $6$\,Gyr, whereas GAMA~569709 shows weak nebular emission. HIPASS~J1159-19~S2 is dominated by young stellar populations associated with active starburst \ion{H}{ii} regions, with comparatively little contribution from an old underlying component.

    \item 
    Gas-phase metallicities derived from star-forming spaxels yield $12+\log(\mathrm{O/H}) \simeq 8.0$-$8.1$ for GAMA~79098 and HIPASS~J1159-19~S2, consistent with their sub-solar stellar metallicities and low stellar masses. GAMA~569709 lacks sufficiently strong emission lines for a reliable study of the nebular line, which implies that it is a gas-depleted or recently quenched dwarf.

    \item 
    Spatially resolved BPT diagnostics indicate that ionisation in GAMA~79098 and HIPASS~J1159-19~S2 is predominantly driven by star formation. Although the [\ion{S}{ii}]-based diagnostic places GAMA~79098 close to the AGN boundary, the discrepancy with the [\ion{N}{ii}] diagnostic, together with the absence of supporting X-ray and mid-infrared signatures, suggests that the observed line ratios are more plausibly explained by low-metallicity effects, hard stellar radiation fields, or shocks rather than AGN activity.

    \item 
    GAMA~79098 and HIPASS~J1159-19~S2 exhibit coherent rotational components in their H$\alpha$ velocity fields, with rotation amplitudes comparable to their measured velocity shear. In contrast, GAMA~569709 shows no clear evidence of ordered rotation within the region probed by detectable H$\alpha$ emission.

    \item 
    The two GAMA group dwarfs are broadly consistent with the LG MZR within uncertainties, although they may suggest marginally higher stellar metallicities at fixed mass; however, the limited sample size precludes firm conclusions. In contrast, HIPASS~J1159-19~S2 lies significantly above the LG relation for its stellar mass, indicating enhanced metallicity relative to comparable LG dwarfs. This offset may reflect differences in chemical enrichment histories, as the HIPASS group is likely in a younger stage of evolution.
\end{enumerate}

\subsection*{Comparison with Local Group Analogues}
A central objective of this study and the Delegate Survey is to compare the observed LGA dwarfs with analogous LG dwarfs. GAMA~79098 closely resembles the SMC in luminosity, stellar mass, metallicity, rotational support, and extended star formation history, and can be regarded as a lower-mass SMC analogue, although located at a substantially larger projected distance from its host galaxy. In contrast, GAMA~569709 shows properties consistent with transition-type or early-type LG dwarfs such as NGC~147 and NGC~185, including comparable stellar mass and metallicity, a dominant old stellar population, weak present-day star formation, and little evidence for strong ordered rotation. Taken together, the structural, chemical, and kinematic properties of the GAMA LGA dwarfs broadly overlap with LG systems, at the level of this pilot sample. HIPASS~J1159-19~S2 exhibits characteristics intermediate between the SMC and LMC, with comparable kinematics and metallicity but dominated by very young stellar populations, suggesting that it is at an earlier stage of evolution. Its proximity to its host galaxy and location within a dynamically young, gas-rich group suggest that environmental processes may play a more significant role than in many LG counterparts.

\subsection*{Implications for the Delegate Survey and Future Work}

This study represents the first resolved spectroscopic analysis within the Delegate Survey and demonstrates that KOALA+AAOmega observations provide sufficient spatial and spectral fidelity to characterise stellar populations, nebular abundances, and internal kinematics of dwarf galaxies at distances of $\sim20$--$30$\,Mpc. To date, $44$ nights of observing time have been allocated to the project, with completed observations for $15$ dwarf galaxies pending analysis. Expansion to this larger sample will enable statistically robust constraints on scaling relations such as the mass-metallicity relation and the Tully-Fisher relation in LGAs. In parallel, comparison of the Delegate LGA sample with mock IFU observations of Local Group dwarfs and with $\Lambda$CDM cosmological simulations will allow a further assessment of whether the LG represents a typical galaxy group in the cosmological context.

Future technical development will focus on refinement of the reduction pipeline, including improved multi-night frame combination to maximise SNR and spatial uniformity, enhanced throughput correction and sky subtraction, as well as completion of the \textsc{PyKOALA} reduction workflow for the $9200$\,\AA\ configuration to access additional near-infrared diagnostics. Continued improvement of fitting procedures, including approaches less dependent on high continuum SNR than the current \textsc{pPXF} based algorithm, will be essential for extending the analysis to fainter, lower surface-brightness dwarfs. Moreover, we will also model the dynamics of Delegate dwarfs to estimate their dynamical masses, constrain their dark matter fractions, and, where possible, resolve their mass profiles and determine their positions on the radial acceleration relation.

\section*{Acknowledgements}
The authors acknowledge Andrew Battisti for guidance on dust extinction corrections. Ethan Crosby is thanked for his support and guidance on using \textsc{pPXF} pipelines. We thank Marcel Pawlowski for his careful and constructive review, which helped improve the clarity and quality of this manuscript.

This research is part of the Delegate Survey collaboration. We acknowledge the contributions of the entire Delegate Survey team in the development of the observational programme, data reduction workflows, and scientific discussions that made this work possible. The Delegate survey is based on data acquired at the Anglo-Australian Telescope, under program A/2024B/10, R/2024B/08, A/2024B/16, R/2025B/06, A/2025A/01, R/2025A/11, A/2025A/01, A/2025A/07, A/2025B/09, A/2025B/28, A/2026A/03, A/2026A/03. We acknowledge the traditional custodians of the land on which the AAT stands, the Gamilaraay people, and pay our respects to elders past and present. 

S.M.S. acknowledges funding from the Australian Research Council (DE220100003).

\section*{Data Availability}
The data are not currently available for public access, but will be released through Data Central once the dataset is sufficiently complete and quality-assured.



\bibliographystyle{mnras}
\bibliography{main} 




%



\bsp	
\label{lastpage}
\end{document}